\documentclass[twocolumn,showpacs,preprintnumbers,amsmath,amssymb,superscriptaddress,10pt]{revtex4}
\usepackage{amssymb}
\usepackage{graphicx,color}% Include figure files
\usepackage{bm}% bold math

\begin{document}

\title{Kinetic description of Bose-Einstein condensation with test particle simulations}

\author{Kai Zhou}
\affiliation{Department of Physics, Tsinghua University and Collaborative Innovation
Center of Quantum Matter, Beijing 100084, China}
\affiliation{Institut f$\ddot{u}$r Theoretische Physik, Johann Wolfgang
Goethe-Universit$\ddot{a}$t Frankfurt, Max-von-Laue-Strasse 1, 60438 Frankfurt am Main, Germany}

\author{Zhe Xu \footnote{xuzhe@mail.tsinghua.edu.cn}}
\affiliation{Department of Physics, Tsinghua University and Collaborative Innovation
Center of Quantum Matter, Beijing 100084, China}

\author{Pengfei Zhuang}
\affiliation{Department of Physics, Tsinghua University and Collaborative Innovation
Center of Quantum Matter, Beijing 100084, China}

\author{Carsten Greiner}
\affiliation{Institut f$\ddot{u}$r Theoretische Physik, Johann Wolfgang
Goethe-Universit$\ddot{a}$t Frankfurt, Max-von-Laue-Strasse 1, 60438 Frankfurt am Main, Germany}

\begin{abstract}
We present a kinetic description of Bose-Einstein condensation for particle
systems being out of thermal equilibrium, which may happen for gluons produced 
in the early stage of ultra-relativistic heavy-ion collisions. The dynamics
of bosons towards equilibrium is described by a Boltzmann equation including
Bose factors. To solve the Boltzmann equation with the presence of 
a Bose-Einstein condensate we make further developments of the kinetic 
transport model BAMPS (Boltzmann Approach of MultiParton Scatterings). 
In this work we demonstrate the correct numerical implementations by comparing
the final numerical results to the expected solutions at thermal equilibrium
for systems with and without the presence of Bose-Einstein condensate.
In addition, the onset of the condensation in an over-populated gluon system
is studied in more details. We find that both expected power-law
scalings denoted by the particle and energy cascade are observed in the
calculated gluon distribution function at infrared and intermediate momentum
regions, respectively. Also, the time evolution of the hard scale exhibits 
a power-law scaling in a time window, which indicates that the 
distribution function is approximately self-similar during that time.
\end{abstract}


\maketitle

\section{Introduction}
\label{sec1:intro}
Using ultra-relativistic heavy-ion collisions, deconfined systems of quarks 
and gluons can be explored under extreme conditions of high temperatures and 
high densities. While the thermodynamic properties of quark-gluon systems at 
equilibrium are followed with great interests, the thermalization of
initially non-thermal systems is still an unsolved outstanding issue.

Based on the Color Glass Condensate (CGC) effective field theory
\cite{McLerran:1993ni}, the two colliding nuclei behave as two very dense 
systems of gluons when going to high colliding energies. Meanwhile, 
an intrinsic momentum scale $Q_s$ emerges, below which gluons saturate to 
a density $\sim 1/\alpha_s$ due to the detailed balance between production 
and annihilation processes of gluons in nuclei. After collision of two nuclei,
gluons are freed and evolve to form a so-called glasma 
\cite{Gelis:2010nm,Lappi:2006fp,Weigert:2005us} through a very short 
isotropization stage \cite{Gelis:2013rba,Kurkela:2015qoa}.
Notice that gluons at this time are far from thermal equilibrium. Moreover,
the gluon number density in such glasma can be overwhelmingly higher than
its corresponding thermal equilibrium density with the same energy density.
We denote such system as an over-populated system, which is opposite to an
under-populated system that will thermalize towards the Bose-Einstein
distribution. The authors in Refs. \cite{Blaizot:2011xf,Blaizot:2012qd} have
pointed out that the highly over-populated glasma would coherently enhance
scatterings and give rise to the strongly interacting nature even though
the coupling is weak. More important, on the way of thermalization the excess
of gluons might be removed into a Bose-Einstein condensate (BEC). The
dynamics of the condensation and thermalization is the main concern of this work.

A BEC is a macroscopic occupation in the ground state, where the de Broglie 
wavelength of particles is larger than the inter-particle scale and thus
particles overlap to form a coherent state. The formation of a BEC is
a fundamental consequence of quantum statistics. Above a certain critical 
density or below a certain critical temperature any further added bosons
must enter into the ground state. The over-populated initial conditions of 
gluons in ultra-relativistic heavy-ion collisions may also lead to the 
formation of a gluon condensate, if the number conserving processes,
i.e. elastic collisions, dominate (at least at early times) the kinetic 
evolution of gluons.

Many studies have been devoted to the non-equilibrium dynamics within either
kinetic approach or classical field theory. In kinetic approach the role of
binary collisions has been investigated first.
In Refs. \cite{Blaizot:2013lga,Blaizot:2014sha,Huang:2013lia} 
under the assumption of small angle scatterings the Boltzmann equation is
approximated by the Fokker-Planck diffusion equation, with which the 
momentum distribution function of the over-populated system of massless gluons
is evolved towards the onset of the condensation. Within the same framework the
quark degrees of freedom are also taken into account in \cite{Blaizot:2014jna},
in order to study the effect on the BEC formation. The Fokker-Planck equation 
is further extended to include mass effects for gauge bosons 
\cite{Blaizot:2015wga} and thus can be used to go beyond the onset and to 
study the gluon condensation and thermalization \cite{Blaizot:2015xga}. 
The situation becomes more complicated, when including number-changing 
processes. On the one hand, number-changing processes will destroy any BEC 
at a long time scale, on the other hand, how they affect the BEC formation 
at a short time scale is still under debates, see 
Refs. \cite{Huang:2014iwa,Huang:2013lia, Blaizot:2016iir}.

The classical-statistical real time lattice simulation for classical 
Yang-Mills field 
\cite{Berges:2013fga,Berges:2013eia,Kurkela:2012hp,Schenke:2016ksl}
is another way to study the non-equilibrium dynamics.
It is shown in \cite{Berges:2012us, Berges:2010ez, Berges:2013fga}
that a compatible agreement between the lattice simulation
and a vertex-resumed effective kinetic approach is achieved.
In classical-statistical real time lattice simulations number changing 
processes are naturally included and a transient condensate emerges at some 
intermediate stage. This is similar as the experimentally observed BEC
for photons in an optical microcavity \cite{photon}. However,
we would like to point out that the existence of BEC in non-Abelian gauge 
theories is actually still under debates, see 
Refs. \cite{Kurkela:2014tea,Kurkela:2011ti}, thereby the thermalization
of a weakly coupled non-Abelian plasma is investigated numerically by solving
the effective kinetic equation, and no condensation happens
in the evolution.

BEC formation has been widely investigated in other fields. In the context of 
early universe cosmology, similar issue about BEC formations has been 
discussed for massive scalar field \cite{Semikoz:1994zp}, whereby the full 
Boltzmann equation is solved for two different systems, one for scalar bosons
under weakly $\lambda \phi^4$ self-interactions and another one for an ideal
Bose gas coupled to a cold fermion gas. In both cases self-similar solutions
in power-law forms are found during the onset of the condensation, which
exhibited turbulent type cascades. Also for cold atom systems the BEC formation
has been studied \cite{Lacaze:2001qf,Pantel:2012pi} in a non-relativistic
regime. In Refs. \cite{pomeau,jackson} both the kinetic equation for the gas
and Gross-Pitaevskii equation for the condensate are combined to describe
the non-equilibrium dynamics of dilute systems of weakly interacting bosons
with the presence of a BEC.
Recently, similar studies for massive scalar bosonic systems with the
relativistic dispersion relation have been done by solving the full
Boltzmann equation numerically in a static case \cite{Meistrenko:2015mda} and
in the case with a longitudinal expansion \cite{Epelbaum:2015vxa}.
The dynamics of BEC formation far from equilibrium has also been studied by 
employing classical-statistical real time lattice simulations for scalar 
field with and without expansion
\cite{Berges:2012us, Berges:2010ez, Dusling:2010rm,Epelbaum:2011pc,Berges:2015ixa}.
Such a treatment has also been applied to describe the cold axion dark 
matter in the universe \cite{Berges:2014xea}.

Most of kinetic approaches that describe Bose-Einstein condensation solve the 
Boltzmann equation in momentum space on a lattice. Different from these 
approaches we have employed a transport parton cascade BAMPS (Boltzmann 
Approach of MultiParton Scatterings) \cite{Xu:2004mz}, which solves the 
Boltzmann equation in full phase space with test particle simulations.
With an improved version of BAMPS we have demonstrated in \cite{Xu:2014ega},
for the first time, the thermalization of gluons with a dynamic Bose-Einstein
condensation in a static system. The present work is a further extension of 
\cite{Xu:2014ega} with all numerical details on the implementations of 
collisions of bosons and the growth of the condensate in BAMPS.
The numerical implementations that we will present can be applied to any
perturbative interactions and are more efficient than those given e.g. 
in \cite{Scardina:2014gxa}. To prove the numerical implementations we
consider a simple case of a non-expanding system with an isotropic momentum
distribution. We assume the dominance of elastic scatterings and ignore
number-changing processes. The role of the latter will be presented in
a forthcoming paper. Besides the checks of the numerical implementations
we will focus on the turbulent cascades and self-similarity of the solution
of the Boltzmann equation for an over-populated system before condensation.

The paper is organized as follows. In Sec. \ref{sec2:bose} Bose statistical
factors are included in BAMPS and a new stochastic method for simulating 
collisions of bosons  is presented. The numerical implementations are proven
for two kind of initial conditions. The one is assumed to be at equilibrium.
Collision rates at various temperatures are evaluated from BAMPS calculations
and compared with the expected analytical values. The another initial condition
is an out of equilibrium state, with which we check its thermalization
towards the Bose-Einstein distribution with the expected temperatures and
chemical potentials. After checking the numerical implementations we then
consider an out of equilibrium and over-populated gluon system and study
first the evolution of the system towards the onset of the Bose-Einstein
condensation in Sec. \ref{sec3:onset}. We will show the appearance of the
turbulence cascades when looking at the time evolution of the momentum
distribution. In Sec. \ref{sec4:bec} we continue the calculation performed
in Sec. \ref{sec3:onset} and evolve the system beyond the onset towards the
full equilibrium with a complete Bose-Einstein condensation. In order to
solve the Boltzmann equation with growing BEC, we first derive a constraint
on the matrix element of scatterings involving massless condensate particles
and then implement such scatterings in BAMPS numerically. To demonstrate 
the correct numerical implementation, the final momentum distribution from
the BAMPS calculation is compared with the expected analytical one.
By looking at the time evolution of the hard momentum scale of the system we
realize the self-similarity of the momentum distribution function within a
certain time window. We will summarize in Sec. \ref{sec5:sum}.
Details on the derivations of some equations and checks on the numerical 
approximations are given in Appendixes. 

\section{Bose statistics in transport calculations}
\label{sec2:bose}
In this section we consider a spatially homogeneous and static boson system,
which momentum distribution at equilibrium is the Bose-Einstein distribution
\begin{equation}
\label{equil}
f_{eq}({\mathbf p}) =\frac{1}{e^{(E-\mu)/T}-1}
\end{equation}
with the temperature $T$ and chemical potential $\mu$. We assume that
particles are massless, while the generalization of the following presented
algorithm for massive particles is straightforward. In Eq. (\ref{equil})
we have then $E=p=|{\mathbf p}|$.
The kinetic equation governing the time evolution of $f$ is the Boltzmann
equation including Bose statistics,
\begin{eqnarray}
\label{boltzmann}
&&\left ( \frac{\partial}{\partial t} + \frac{{\mathbf p}_1}{E_1}
\frac{\partial}{\partial {\mathbf r}} \right )\, 
f_1 = \frac{1}{2E_1} \int d\Gamma_2 \frac{1}{2} \int d\Gamma_3 d\Gamma_4
| {\cal M}_{34\to 12} |^2 \nonumber \\
&& \times \ \left[ f_3 f_4 (1+f_1) (1+f_2) - f_1 f_2 (1+f_3) (1+f_4) \right] 
\nonumber \\
&& \times (2\pi)^4 \delta^{(4)} (p_3+p_4-p_1-p_2) \,,
\end{eqnarray}
where $f_i=f_i({\mathbf r}, {\mathbf p}_i, t)$ and 
$d\Gamma_i=d^3p_i/(2E_i)/(2\pi)^3, i=1,2,3,4$. Binary collisions 
$34\to12$ and $12\to34$ are determined by the collision kernel
$| {\cal M}_{34\to 12} |^2$ and  $| {\cal M}_{12\to 34} |^2$, 
which are equal. 
$(1+f_1)(1+f_2)$ and $(1+f_3)(1+f_4)$ are the Bose factors, with which
the distribution (\ref{equil}) is a solution of Eq. (\ref{boltzmann}).

The phase space distribution function $f$ is represented by test particles
with position and momentum. A test particle moves along its classical
trajectory and will change the direction, once it collides with other test
particles. Free streaming and collision are two components of test particle
simulations for solving the Boltzmann equation.
In this section we concentrate on the numerical implementation of 
collisions of bosons with Bose statistics.

For two particles with momenta in the range
(${\bf p}_3, {\bf p}_3+\Delta{\bf p}_3$) and 
(${\bf p}_4, {\bf p}_4+\Delta{\bf p}_4$), and in the same spatial volume
element $\Delta^3 x$, the collision rate per unit phase space for such
particle pair can be read off from the collision term in Eq. (\ref{boltzmann}),
\begin{eqnarray}
\label{collrate}
\frac{\Delta N_{coll}}{\Delta t \Delta^3 x \Delta^3 p_3} &=& 
\frac{1}{(2\pi)^3 2E_3} \frac{\Delta^3 p_4}{(2\pi)^3 2E_4} f_3 f_4 \nonumber \\
&& \times \frac{1}{2} \int d\Gamma_1 d\Gamma_2
| {\cal M}_{34\to 12} |^2  (1+f_1)(1+f_2) \nonumber \\
&& \times (2\pi)^4
\delta^{(4)} (p_3+p_4-p_1-p_2) \,.
\end{eqnarray}
Similar to the usual definition of the cross section
(for massless identical particles)
\begin{equation}
\label{cs22}
\sigma_{22}= \frac{1}{4s} \int d\Gamma_1 d\Gamma_2
| {\cal M}_{34\to 12} |^2 (2\pi)^4 \delta^{(4)} (p_3+p_4-p_1-p_2) \,,
\end{equation}
we define an effective cross section involving the Bose factors
\begin{eqnarray}
\label{cs22eff}
\sigma_{22}^{eff}&=& \frac{1}{4s} \int d\Gamma_1 d\Gamma_2
| {\cal M}_{34\to 12} |^2 (1+f_1)(1+f_2) \nonumber \\
&& \times (2\pi)^4 \delta^{(4)} (p_3+p_4-p_1-p_2) \,,
\end{eqnarray}
where $s$ is the invariant mass of the particle pair.
When expressing the phase space distribution functions as
\begin{equation}
\label{distf}
f_i=\frac{\Delta N_i}{N_{test} \frac{1}{(2\pi)^3} \Delta^3 x \Delta^3 p_i},
\quad i=3,4,
\end{equation}
where $\Delta N_i$ are the particle numbers counted in the phase space 
elements, one obtains the collision probability in a volume element 
$\Delta^3 x$ within a time interval $\Delta t$
\begin{equation}
\label{p22}
P_{22} = \frac{\Delta N_{coll}}{\Delta N_3 \Delta N_4} = 
v_{rel} \frac{\sigma_{22}^{eff}}{N_{test}} \frac{\Delta t}{\Delta^3 x}\,,
\end{equation}
$v_{rel}=s/2E_3E_4$ denotes the relative velocity of two incoming particles
and $N_{test}$ is the number of test particles per real particle.

With $P_{22}$ one can simulate a collision of two particles in a stochastic
way by using Monte Carlo technique, as introduced in \cite{Xu:2004mz}:
One samples a random number between $0$ and $1$. A collision occurs, if 
this number is smaller than $P_{22}$. In this case, momenta of particles
are changed according to the distribution of the collision angle.
This so-called standard stochastic method has been employed to simulate 
collisions of bosons \cite{Scardina:2014gxa}.
However, to obtain $\sigma_{22}^{eff}$ for each particle pair,
one has to carry out integrals numerically, which demands a huge 
computing power, although the integral in Eq. (\ref{cs22eff}) can be reduced 
to a two dimensional integration over the solid collision angle in the center
of mass frame of the two colliding particles, $\Omega^*$.
To avoid this disadvantage we introduce a new scheme, which has been
employed in our previous work \cite{Xu:2014ega}. Here we give more details.
Instead of the collision probability we define a differential collision probability
\begin{equation}
\label{dp22}
\frac{dP_{22}}{d\Omega^*}=\frac{v_{rel}}{N_{test}} 
\frac{d\sigma_{22}}{d\Omega^*}
(1+f_1) (1+f_2) \frac{\Delta t}{\Delta V} \,.
\end{equation}
The integration over $\Omega^*$ gives the total collision probability
in Eq. (\ref{p22}). In contrast to the standard stochastic method, we
introduce a new stochastic method inspired from the Monte Carlo integration
over $\Omega^*$. First we choose a reference distribution function
$dF/d\Omega^*$, which is normalized to $1$. Second, a solid angle 
$\tilde \Omega^*$ is sampled according to $dF/d\Omega^*$ for each particle
pairs. With the momenta of incoming particles, ${\mathbf p}_3$
and ${\mathbf p}_4$, and the solid collision angle $\tilde \Omega^*$ we
can determine the momenta of outgoing particles, ${\mathbf p}_1$ and
${\mathbf p}_2$, and, thus, determine the Bose factor $(1+f_1)(1+f_2)$
from the extracted $f$ at ${\mathbf p}_1$ and ${\mathbf p}_2$, respectively.
Third, we sample a random number between zero and the value of $dF/d\Omega^*$
at $\tilde \Omega^*$. A collision occurs, if this random number is smaller
than $dP_{22}/d\Omega^*$ at $\tilde \Omega^*$. 

The advantage of the new scheme is that we do not need to calculate
$\sigma_{22}^{eff}$. On the other hand, we have to sample the momenta
of outgoing particles to obtain the Bose factor, before we decide whether
a collision actually occurs. This costs an extra computing time, but is much
less time consuming than the integration for $\sigma_{22}^{eff}$. 

We note that in the standard rejection method the reference function is always
larger than the distribution function. In our case, it is not ensured that
$dF/d\Omega^*$ is always larger than $dP_{22}/d\Omega^*$. For instance,
for a Bose-Einstein distribution [$\mu=0$ in Eq. (\ref{equil})] the Bose
factor $(1+f_1)(1+f_2)$ could be infinite, when $p_1$ or $p_2$ approaches
$0$. If the sampled $\tilde \Omega^*$ lies in the region, where 
$dP_{22}/d\Omega^*$ is larger than $dF/d\Omega^*$, all the previous
operations done within the current time step should be redone with
a smaller time step, which reduces $dP_{22}/d\Omega^*$ [see Eq. (\ref{dp22})]
to be smaller than $dF/d\Omega^*$.

Moreover, although $dF/d\Omega^*$ can be chosen arbitrarily,
the sampling will become more efficient, if the shape of $dF/d\Omega^*$ 
is more similar to $dP_{22}/d\Omega^*$. In practice, we choose 
$dF/d\Omega^*=(d\sigma_{22}/d\Omega^*)/\sigma_{22}$, if $\sigma_{22}$ can be
obtained analytically. Thus, to decide whether a collision occurs, one
only needs to compare
\begin{equation}
\label{p22p}
P'_{22}=v_{rel}\frac{\sigma_{22}}{N_{test}} \frac{\Delta t}{\Delta V}
(1+f_1) (1+f_2)
\end{equation}
with a random number between $0$ and $1$.

The Bose factor $(1+f_1)(1+f_2)$ is essential for the dynamics of bosons at low
momentum when $f_1$ and/or $f_2$ is larger than $1$. Therefore, a precise 
extraction of $f$ at low momentum is quite important.
Since $f$ is the particle density in phase space, which has six dimensions,
we shall use large number of $N_{test}$, in order to reduce statistical
fluctuations and to obtain precise values of the Bose factor $(1+f_1)(1+f_2)$.
In this work we assume for simplicity that $f$ is homogeneous in coordinate
space and is isotropic in momentum space. Thus, $f$ depends on $p$ only.
We extract $f$ at equidistant $p_i, i=0,1,2,\cdots$, beginning at 
$p_0=5 \mbox{ MeV}$ and separated by an interval of 
$\Delta p=2.5 \mbox{ MeV}$. The value of $f$ at $p_i$ is obtained by the 
number of test particles within the interval $[p_i-\Delta p/2:p_i+\Delta p/2]$.
The value of $f$ at $p > p_1=7.5 \mbox{ MeV}$ and $p \neq p_i$
is obtained by interpolation, while the value of $f$ at $p < p_1$ is obtained 
by extrapolation using a power law function, which fits $f$s at first ten
$p_i$ beginning from $p_1$.

To prove the new stochastic method presented above, we perform numerical 
calculations for massless bosons in a static cubic box. The size of the box
is set to be $3 \times 3 \times 3 \mbox{ fm}$. We use a periodic boundary
condition to cancel the expansion. The box is divided into cubic cells
with equal volume $\Delta V$. The cell length is set to be $0.125 \mbox{ fm}$.
For the demonstration we consider binary collisions with a constant total 
cross section of  $\sigma_{22}=10 \mbox{ mb}$
and an isotropic distribution of the collision angle, which corresponds to
 $|{\cal M}|^2=32\pi s \sigma_{22}$. In addition, bosons are assumed to
have a degeneracy factor $g=1$.

\begin{figure}[t]
\centering
\includegraphics[width=0.45\textwidth]{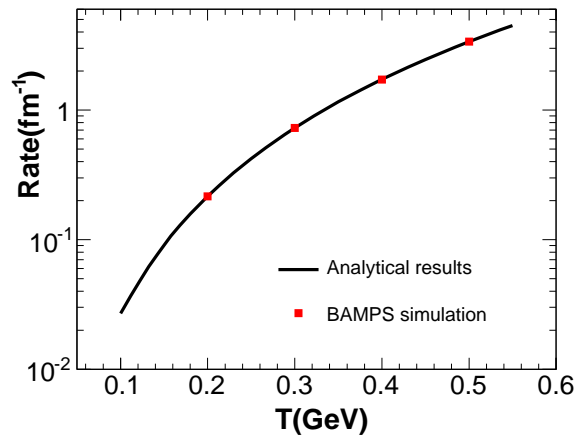}
\caption{Collision rates per particle as a function of temperature. 
The solid squares show the numerically calculated collision rates with 
$N_{test}=1600$, while the solid curve depicts the analytical results.}
\label{fig1-ratecom}
\end{figure}
For the first test we assume an equilibrium initial condition obeying the 
Bose-Einstein distribution, i.e., $\mu=0$ in Eq. (\ref{equil}). We compare 
the collision rate per particle obtained from the numerical calculations with 
the analytical results by integrating Eq. (\ref{collrate}) over the full 
phase space. Figure \ref{fig1-ratecom} shows the comparisons for various
temperatures. The squares denote the numerical results and the solid curve 
depicts the analytical rates. We see a very good agreement between the 
numerical and analytical collision rates.

\begin{figure}[b]
\centering
\includegraphics[width=0.45\textwidth]{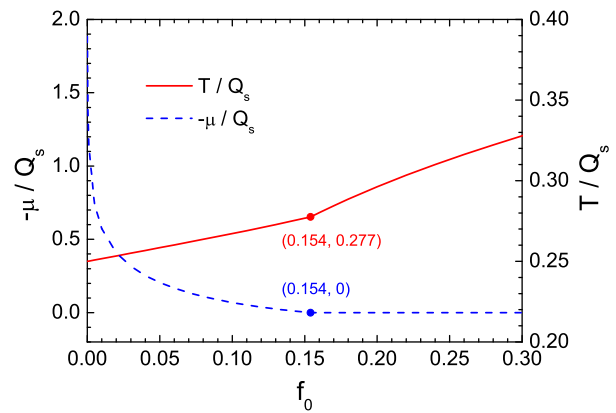}
  \caption{Dependence of the chemical potential $\mu$ and the temperature $T$
on $f_0$ and $Q_s$.}
\label{fig2-eqf0}
\end{figure}
As the next we prove the equilibration of bosons with an out of equilibrium
initial distribution,
\begin{equation}
\label{init}
f_{init}({\mathbf p})=f_0 \theta (Q_s-|{\mathbf p}|) \,,
\end{equation}
which resembles that in the early stage of ultrarelativistic heavy ion
collisions \cite{Blaizot:2011xf}. $f_0$ and $Q_s$ are parameters, which
simply model the relation to the colliding energy. The higher the colliding
energy, the larger are $f_0$ and $Q_s$, and thus the larger are the particle
number and energy density, which can be obtained from Eq. (\ref{init})
\begin{equation}
n_{init}=g f_0 \frac{Q_s^3}{6 \pi^2} \,, \ \
e_{init}=g f_0 \frac{Q_s^4}{8 \pi^2} \,,
\end{equation}
where $g=1$ in this section. 
The initial momentum distribution has been simplified to be isotropic, although
the new method introduced above can be applied for anisotropic momentum
distributions.

For calculations in a box the energy density $e$ is conserved. Assuming binary
collisions of particles only, the particle number density $n$ is also conserved
during the equilibration. In particular, $n$ and $e$ are equal to those at
equilibrium with the distribution function (\ref{equil}). Therefore, for given
$f_0$ and $Q_s$ from the initial condition, the temperature $T$ and the
chemical potential $\mu$ at equilibrium can be calculated. 
In Fig. \ref{fig2-eqf0} we plot $T/Q_s$ and $-\mu/Q_s$ as functions of $f_0$.
The kinks at $f_0^c=0.154$ indicate a transition from normal gluon gas to 
the one with the appearance of a BEC, since $\mu$ must keep zero for
increasing $f_0$ (or density).

For $f_0>f^c_0$ the particle system is initially over-populated and a BEC
will occur during the equilibration. At thermal equilibrium the distribution
function contains the Bose-Einstein distribution and a condensate,
\begin{equation}
\label{equil_BEC}
f_{eq}({\mathbf p})=\frac{1}{e^{E/T}-1}+
(2\pi)^3 n_c^{eq} \delta^{(3)}({\mathbf p}) \,,
\end{equation}
where $n_c^{eq}$ is the density of the condensate particles. From the particle
number and energy conservation we obtain easily that
\begin{eqnarray}
\label{temp}
T&=& \left ( \frac{15f_0}{4} \right )^{1/4} \frac{Q_s}{\pi} \,, \\
\label{nceq}
n_c^{eq} &=& n_{init} \left [ 1- \zeta(3) \frac{6}{\pi^3} \left ( \frac{15}{4}
\right )^{3/4}  \left ( \frac{1}{f_0} \right )^{1/4} \right ] \,.
\end{eqnarray}
From the above equation we can also obtain $f_0^c$, at which $n_c^{eq}=0$.
We see that $f_0^c$ is independent of $Q_s$.

For $f_0 < f_0^c$ the system is under-populated. In this section we present
thermalization of systems with $Q_s=1 \mbox{ GeV}$ and two sets of $f_0$,
$f_0=0.05$ and $f_0=f_0^c$. In both cases no BEC will appear.
The difference from one to another case is that for $f_0=0.05$ the equilibrium
momentum distribution $f_{eq}$ converges to $1/(e^{-\mu/T}-1)$ at $p\to 0$,
while for $f_0=f_0^c$ it diverges at $p \to 0$.

\begin{figure}[ht]
\includegraphics[width=0.45\textwidth]{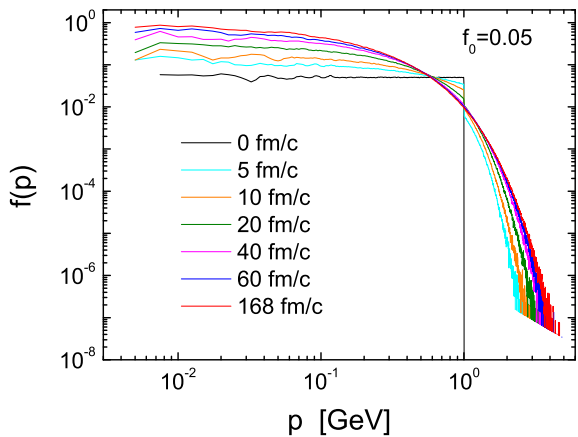}
  \caption{The time evolution of the momentum distribution function for
$f_0=0.05$.}
\label{fig3-005f}
\end{figure}

For $f_0=0.05$ the time evolution of the particle momentum distribution
is shown in Fig. \ref{fig3-005f}. 
$f$ is calculated at a set of equidistant 
momenta with $\Delta p=2.5 \mbox{ MeV}$, beginning from $5 \mbox{ MeV}$.
We find a gradual equilibration with a large 
timescale of about $170 \mbox{ fm/c}$. During the equilibration particles,
which mainly populate at $Q_s$ in the initial condition, flow into the
lower and higher momentum region. We see that the statistical
fluctuation is strong at very low and very high momentum region due to small
particle populations. One needs large value of $N_{test}$ to reduce these
fluctuations. In this calculation $N_{test}=691000$ is used.

\begin{figure}[ht]
\includegraphics[width=0.45\textwidth]{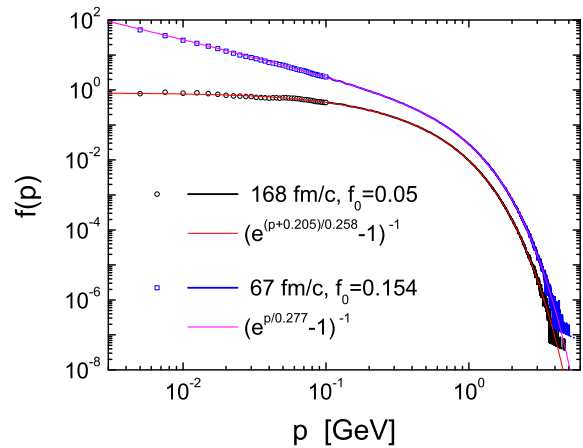}
  \caption{Comparisons between the calculated distributions
at equilibrium times and the Bose-Einstein distributions with the expected
temperatures and chemical potentials.}
\label{fig4-be1}
\end{figure}

From the calculations, which results are presented in Fig. \ref{fig2-eqf0},
we obtain $T=0.258 \mbox{ GeV}$ and $\mu=-0.205 \mbox{ GeV}$ for
$f_0=0.05$. In Fig. \ref{fig4-be1} we compare the momentum distribution
at $168 \mbox{ fm/c}$ with the thermal equilibrium distribution (\ref{equil}).
The open circles depict the first $40$ values of the calculated distribution
separated by a equidistant interval of $\Delta p=2.5 \mbox{ MeV}$ and
beginning with $5 \mbox{ MeV}$. We see a perfect agreement over $7$ orders
in magnitude.

\begin{figure}[b]
\includegraphics[width=0.45\textwidth]{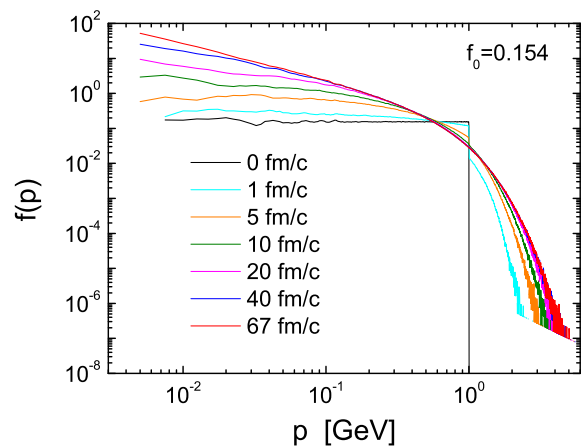}
  \caption{Same as Fig. \ref{fig3-005f}, but for $f_0=0.154$.}
\label{fig5-crif}
\end{figure}
The equilibration for $f_0=f_0^c$ is presented in Fig. \ref{fig5-crif}.
In this calculation $N_{test}=230000$ is used. We see the divergence of
$f(p)$ at $p\to 0$. The equilibration has a shorter timescale of about
$65 \mbox{ fm/c}$ due to a larger density, compared with that for $f_0=0.05$. 
Also, the calculated momentum distribution at $67 \mbox{ fm/c}$
agrees well with the Bose-Einstein distribution with 
$T=0.277 \mbox{ GeV}$ and $\mu=0$, as seen in Fig. \ref{fig4-be1}.

The perfect agreements between the numerical results and analytical
solutions shown in this section demonstrate the correct implementation
of the new method solving the Boltzmann equation with Bose statistics.

\section{The onset of Bose-Einstein condensation}
\label{sec3:onset}
After we have proven the numerical implementation for collisions of bosons
in the previous section, we consider in the rest of the paper an over-populated
system of massless gluons. For this we set $Q_s=1 \mbox{ GeV}$ and $f_0=1$ in
the initial distribution (\ref{init}). For the BAMPS calculation $N_{test}=2400$
is used. Although $N_{test}$ is smaller than those used for $f_0=0.05$ and
$f_0=0.154$ in the previous section, the total number of test particles is
almost the same, because gluons have a degeneracy factor of $g=16$. In this
section we study the onset of Bose-Einstein condensation.
In the next section we demonstrate the full thermalization of gluons with 
a complete Bose-Einstein condensation. 

The elastic scatterings of massless gluons are described in leading order of
perturbative QCD. We use the same matrix element as that in our previous 
work \cite{Xu:2014ega},
\begin{equation}
\label{matrix}
| {\cal M}_{gg\to gg} |^2 \approx 144 \pi^2 \alpha_s^2 \frac{s^2}{t (t-m^2_D)} \,,
\end{equation}
which is calculated by using the Hard-Thermal-Loop (HTL) treatment
\cite{Aurenche:2002pd,Kurkela:2011ti}.
$s$ and $t$ are the Mandelstam variables, and $m_D$ is the screening mass
\begin{equation}
\label{md2}
m^2_D=16\pi N_c \alpha_s \int \frac{d^3p}{(2\pi)^3} \, \frac{1}{p} f \,.
\end{equation}
The matrix element obeys the general condition for the occurrence of 
Bose-Einstein condensation. This will become clear in the next section,
when we describe the condensation of gluons. The coupling is set to be 
$\alpha_s=0.3$ throughout the paper. 

According to the definition (\ref{cs22}) we obtain the total cross section
\begin{equation}
\label{cs22no}
\sigma_{gg\to gg}=\frac{9}{2} \pi \frac{\alpha_s^2}{m_D^2} 
\ln \frac{1-m_D^2/t_{cut}}{1+m_D^2/s}
\,,
\end{equation}
where the logarithmic divergence has been regularized by an upper cutoff of 
$t$, $t_{cut}$. $t_{cut}$ is determined in consistency with the cross section
of collisions involving condensate particles, which will be clarified in the
next section. We note that scatterings with $t$ approaching to zero do not
contribute to thermalization.

\begin{figure}[ht]
 \centering
 \includegraphics[width=0.45\textwidth]{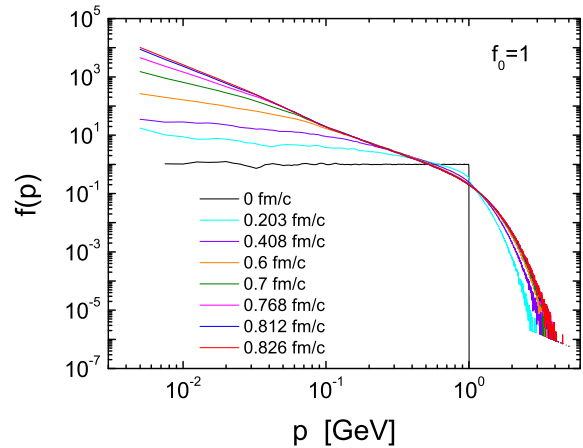}
 \caption{Same as Fig. \ref{fig3-005f}, but for $f_0=1$.}
 \label{fig6-dist}
\end{figure}

Figure \ref{fig6-dist} shows the time evolution of the distribution function.
At the first sight we see a transportation of particles as well as energies 
from $Q_s=1 \mbox{ GeV}$ towards regions of lower and higher momentum. 
More elaborate analyses carried out below will expose finer structures of 
$f(p)$.

Theoretical studies \cite{Semikoz:1994zp} showed that the kinetic Boltzmann 
equation (\ref{boltzmann}) has temporally self-similar solutions. 
In a quasi-stationary state  $f(p)$ is a scaling invariant power law function 
$f(p) \sim p^{-r}$ at momentum $p$, where $f(p) \gg 1$. The exponent $r$
depends on the scaling behavior of the matrix element under rescaling of the
momentum. For our case (\ref{matrix}) we follow the derivations in 
Ref. \cite{Semikoz:1994zp,Meistrenko:2015mda} 
and obtain $r=2$ in a turbulent state with constant transport of particle 
number (called as particle cascade), and $r=7/3$ in a turbulent
state with constant energy transport (called as energy cascade).

\begin{figure}[ht]
 \centering
 \includegraphics[width=0.45\textwidth]{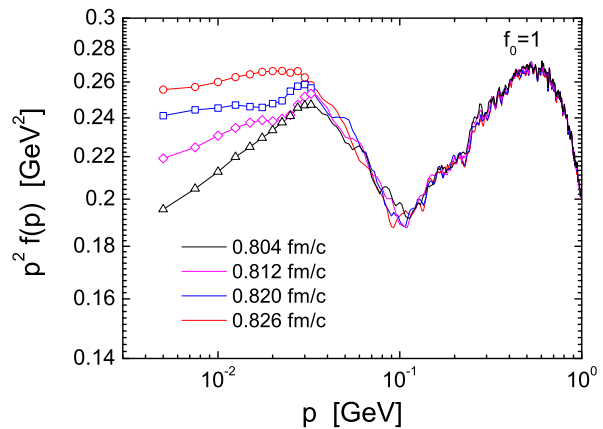}
 \caption{The particle distribution function multiplied by momentum squared.}
 \label{fig7-distp2}
\end{figure}

We show $p^2 f(p)$ at four various times in Fig. \ref{fig7-distp2}.
The symbols depict the extracted values of $f$ at first $12$ momenta. 
If there exists a particle cascade [$f(p) \sim p^{-2}$], we will see a plateau
of $p^2 f(p)$ within a momentum interval. This is indeed seen from the time 
$0.812$ fm/c to $0.826$ fm/c from $p\approx 0.03 \mbox{ GeV}$ towards the
infrared region.
We note that the extraction of $f(p)$ at $p<5 \mbox{ MeV}$ is inaccurate,
because the number of test particles at the deep infrared region in the present
calculation is too small to overcome statistical fluctuations. Thus,
we end the calculation, when $f(p)$ at $p>5 \mbox{ MeV}$ reaches the scaling 
behavior $f(p) \sim p^{-2}$. With a much larger $N_{test}$ in future 
calculations we could obtain more accurate values of $f(p)$ in the deep 
infrared region.  

\begin{figure}[ht]
 \centering
 \includegraphics[width=0.45\textwidth]{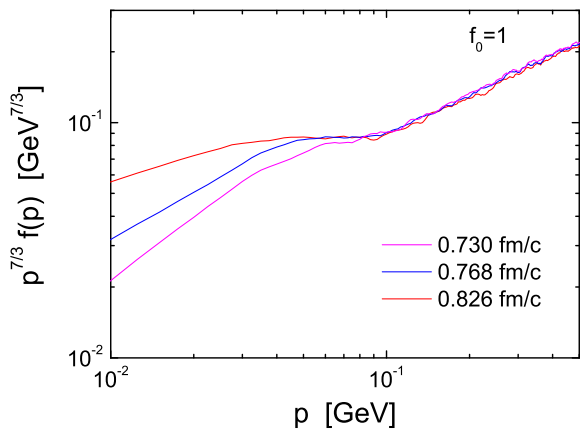}
 \caption{The particle distribution function multiplied by momentum to
 a power of $7/3$.}
 \label{fig8-distp73}
\end{figure}

In Fig. \ref{fig7-distp2} we see that the height of the plateau increases
with time, which indicates that $f(p)$ at the plateau is time dependent and
not stationary. The reason is due to two other power law scalings
at higher momentum. The second power law scaling is between 
$p\approx 0.03 \mbox{ GeV}$ and $0.1 \mbox{ GeV}$, where the exponent is 
larger than $2$, and the next is between $0.1 \mbox{ GeV}$ and  
$0.5 \mbox{ GeV}$, where the exponent is smaller than $2$. Both exponents can 
be extracted through fitting $f(p)$ by using power law functions. We find that
the exponent of the second scaling is $7/3$, which corresponds to the energy 
cascade and is better seen in Fig. \ref{fig8-distp73}, where $p^{7/3} f(p)$
is plotted.
The plateau appears at $p\approx 0.1 \mbox{ GeV}$ at $0.73$ fm/c and extends 
towards lower momentum with increasing time. The $p^{-7/3}$ scaling appears 
earlier than the $p^{-2}$ scaling that appears at about $0.81$ fm/c. 
We also see that the height of the plateau in $p^{7/3} f(p)$,
on the contrary to that in $p^2 f(p)$, does not change, which indicates that
the plateau in $p^{7/3} f(p)$ is the region of the stationary turbulence.
Since the $p^{-2}$ scaling region of $f(p)$ connects the lower momentum end
of the $p^{-7/3}$ scaling region, the extending $p^{-7/3}$ scaling region 
towards lower momentum causes the increase of $f(p)$ in the $p^{-2}$ scaling
region, as already observed in Fig. \ref{fig7-distp2}. 

\begin{figure}[ht]
 \centering
 \includegraphics[width=0.45\textwidth]{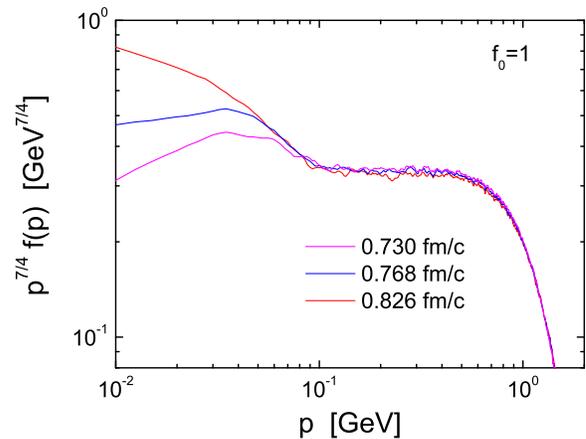}
 \caption{The particle distribution function multiplied by momentum to
 a power of $7/4$.}
 \label{fig9-distp74}
\end{figure}

The stationary turbulence region with $f(p) \sim p^{-7/3}$ is followed by
a further power law scaling with $f(p) \sim p^{-7/4}$, as shown in 
Fig. \ref{fig9-distp74}, where $p^{7/4} f(p)$ is plotted.
This power law scaling is not known so far in the literature. Even not 
obviously, we can recognize that in the new power law scaling region $f(p)$ 
is not stationary. The height of the plateau in Fig. \ref{fig9-distp74} 
decreases slightly with increasing time. Roughly speaking, particles and 
energies in the $p^{-7/4}$ scaling region transport through the stationary 
$p^{-7/3}$ scaling region towards lower momentum region, where the $p^{-2}$ 
scaling region extends to infrared momentum. At the same time it exits 
another particle and energy transport from the $p^{-7/4}$ scaling region
towards higher momentum, which is, however, hardly seen in 
Fig. \ref{fig6-dist}. The observed transportations are due to the initial 
condition given in this study, where most particles and energies are 
initialized at $p=Q_s=1 \mbox { GeV}$.

We note that it is surprising that we see both $p^{-2}$ and $p^{-7/3}$ 
scalings in our calculation, although the theoretical derivations of the two 
power law scalings are done with approximations. More calculations are 
required for the time evolution of $f(p)$ in the infrared momentum region 
and will be done in the future.

Even if future calculations could confirm the $p^{-2}$ power law scaling in 
the deep infrared region, this cannot lead to the formation of a BEC.
To show this, we calculate the particle density at $p=0$ by integrating 
$f(p)$ over a sphere of radius $p_0$ and then going to the limit $p_0 \to 0$,
\begin{equation}
n(p=0)=\lim_{p_0\to 0} \int_0^{p_0} \frac{dp}{2\pi^2} p^2 f(p) \,.
\end{equation}
For $f(p) \sim p^{-2}$ at low momentum, the particle density at $p=0$ vanishes.
The study on the mechanism of a BEC formation is beyond the scope of this
paper. We assume the formation of a small piece of BEC instantaneously
at some timescale. The growth of the BEC can then be described by the kinetic 
Boltzmann equation, which we will present in the next section.

Since the mechanism of the Bose-Einstein condensation of gluons is not known 
yet, it is impossible to determine the exact time when the condensation begins.
Nevertheless, one can make estimates on this timescale by looking at the 
possible onset of the gluon condensation. In our previous work
\cite{Xu:2014ega} we have fitted the gluon distribution at low momenta by the
Bose-Einstein distribution with an effective temperature and an effective 
chemical potential. The latter increases from a negative value to zero, which
is assumed to be reached at the onset of the gluon condensation. 
Thus, we have chosen the moment as the start time of the gluon condensation,
$t_c$, once the effective chemical potential becomes positive due to the 
numerical fluctuation around zero. Thus, at $t_c$ the distribution function 
at low momenta has a $1/p$ power law form. In this section we have found that
the distribution function at low momenta evolves further from $1/p$ to $1/p^2$. 
The latter behavior is expected as a fix point corresponding to a constant
particle transportation. Due to the new observation we choose $t_c$, 
different from that in the previous work, as the time when the $1/p^2$ power 
law distribution is achieved at low momenta ($\sim 5 \mbox{ MeV}$).
The calculation shows that $t_c=0.826 \mbox{ fm/c}$.

\section{Bose-Einstein condensation}
\label{sec4:bec}
In this section we continue the calculate in the previous section and study
the time evolution of the particle distribution 
function $f(p,t)$ in the presence of a BEC. $f(p,t)$ is decomposed into two 
parts $f=f^g+f^c$, where $f^g$ denotes the distribution of gas (noncondensate) 
particles and $f^c=(2\pi)^3 n_c \delta^{(3)}({\mathbf p})$ denotes the
distribution of the condensate particles with zero momentum. Same as for the 
onset of the condensation, we consider here elastic collisions only. Denoting 
gas particles by $g$ and condensate particles by $c$, we consider 
$g+g \to g+g$, $g+c \to g+g$, and $g+g \to g+c$ processes. The Boltzmann 
equations for gas and condensate particles are then given as follows:
\begin{eqnarray}
\label{boltzmann1}
&&\frac{\partial f^g_1}{\partial t} 
= \frac{1}{2E_1} \int d\Gamma_2 \frac{1}{2} \int d\Gamma_3 d\Gamma_4
| {\cal M}_{34\to 12} |^2 \nonumber \\
&& \times \ \left[ f^g_3 f^g_4 (1+f^g_1) (1+f^g_2) 
+f^g_3 f^g_4 (1+f^g_1) f^c_2 \right. \nonumber \\
&& \ \ \ \ + f^c_3 f^g_4 (1+f^g_1) (1+f^g_2)
+f^g_3 f^c_4 (1+f^g_1) (1+f^g_2) \nonumber \\
&& \ \ \ \  - f^g_1 f^g_2 (1+f^g_3) (1+f^g_4) - f^g_1 f^c_2 (1+f^g_3) (1+f^g_4)
\nonumber \\
&& \ \ \ \, \left. - f^g_1 f^g_2 f^c_3 (1+f^g_4) - f^g_1 f^g_2 (1+f^g_3) f^c_4 \right] 
\nonumber \\
&& \times (2\pi)^4 \delta^{(4)} (p_3+p_4-p_1-p_2) \,,
\end{eqnarray}
\begin{eqnarray}
\label{boltzmann2}
&&\frac{\partial f^c_1}{\partial t} = \frac{1}{2E_1} 
\int d\Gamma_2 \frac{1}{2} \int d\Gamma_3 d\Gamma_4
| {\cal M}_{34\to 12} |^2 \nonumber \\
&& \times \ \left[ f^g_3 f^g_4 f^c_1 (1+f^g_2) 
- f^c_1 f^g_2 (1+f^g_3) (1+f^g_4) \right] \nonumber \\
&& \times (2\pi)^4 \delta^{(4)} (p_3+p_4-p_1-p_2) \,.
\end{eqnarray}
The terms of the spatial derivative of $f^g_1$ and $f^c_1$ drop out, since 
we restrict ourselves to consider a spatially homogeneous gluon matter. 

Before we come to the numerical implementations for solving the Boltzmann
equations (\ref{boltzmann1}) and (\ref{boltzmann2}), 
we first show that the matrix element (\ref{matrix}) describes the 
condensation process with a finite rate. For this purpose we
integrate Eq. (\ref{boltzmann2}) over ${\mathbf p}_1$, which gives the time
derivative of the density of condensate particles. After a lengthy calculation
for the integral of the right-hand side of Eq. (\ref{boltzmann2}), which 
details are given in Appendix \ref{appA}, we obtain
\begin{eqnarray}
\label{condrate}
\frac{\partial n_c}{\partial t} = \frac{n_c}{64\pi^3} 
&\int& dE_3 dE_4 \left [ f_3^g f_4^g - f_2^g (1+f_3^g+f_4^g) \right ] 
\nonumber \\
&& \times \ E \left[ \frac{| {\cal M}_{34\to 12} |^2}{s} \right]_{s=2mE} \,.
\end{eqnarray}
The two terms on the right-hand side of Eq. (\ref{condrate}) correspond to
kinetic processes for the condensation and the evaporation, respectively.
$E=E_3+E_4$ is the total energy in the collision, 
$p=|{\mathbf p}_3+{\mathbf p}_4|$
is the total momentum, and $s=E^2-p^2$ is the invariant mass.
$m$ denotes the particle mass at rest, which is zero for gluons. 
From Eq. (\ref{condrate}) we see that in order to describe the gluon 
condensation with a finite rate, the ratio $|{\cal M}_{34\to 12} |^2/s$ at 
$s=0$ should be nonzero and finite. This is fulfilled for the matrix element 
(\ref{matrix}), since
\begin{eqnarray}
\left[ \frac{| {\cal M}_{34\to 12} |^2}{s} \right]_{s=0}&=&
144 \pi^2 \alpha_s^2\left[ \frac{s}{t (t-m^2_D)} \right]_{s=0} \nonumber \\
&=&144 \pi^2 \alpha_s^2 \frac{1}{m^2_D}
\end{eqnarray}
is nonzero and finite. For the constant cross section with the isotropic 
distribution of collision angles the corresponding matrix element is 
proportional to $s$. Therefore, in this case, the condensation process of 
gluons has a finite rate too.

In the following we show how we solve Eqs. (\ref{boltzmann1}) and
(\ref{boltzmann2}) numerically. We note that the numerical implementations are
general and do not require in particular the isotropy of the momentum, which 
is needed to derive Eq. (\ref{condrate}). Since we have assumed the isotropy 
of the momentum for simplicity, Eq. (\ref{condrate}) is automatically solved 
by solving Eq. (\ref{boltzmann2}). From Eq. (\ref{condrate}) we see that 
the growth of the condensate needs an initial density, $n_c(t_c)$. We take 
the density of such gluons, which energy is smaller than $5 \mbox{ MeV}$, 
as $n_c(t_c)$. This numerical handling will become clear later in this section.

Now we present the numerical method simulating collisions. For $g+g \to g+c$, 
which produces a condensate gluon, the integral in the effective cross
section, Eq. (\ref{cs22eff}), can be carried out analytically with help of
delta-functions in $f^c_1 \sim n_c \delta^{(3)}({\mathbf p_1})$ and for 
energy-momentum conservation. That is
\begin{eqnarray}
\label{cs22effc}
\sigma_c^{eff}&=& \frac{1}{2s} \int d\Gamma_1 d\Gamma_2
| {\cal M}_{34\to 12} |^2 f^c_1(1+f^g_2) \nonumber \\
&& \times (2\pi)^4 \delta^{(4)} (p_3+p_4-p_1-p_2) \nonumber \\
&=& \frac{1}{2}  \left[ \frac{| {\cal M}_{34\to 12} |^2}{s} \right]_{s=0}
\int d\Gamma_1 d\Gamma_2 f^c_1(1+f^g_2) \nonumber \\
&& \times (2\pi)^4 \delta^{(4)} (p_3+p_4-p_1-p_2) \nonumber \\
&=& \pi n_c \left[ \frac{| {\cal M}_{34\to 12} |^2}{s} \right]_{s=0}
\frac{1}{2p} [1+f^g(p)] \nonumber \\
&& \times \delta[(E-p)^2] \,.
\end{eqnarray}
Compared with Eq. (\ref{cs22eff}), the factor $1/2$ drops out, because we 
fix particle $1$ to be the condensate gluon. The details of the integration
can be found in Appendix \ref{appB}. 

In principle, one can compute the collision probability $P_{g+g\to g+c}$ by 
putting Eq. (\ref{cs22effc}) into Eq. (\ref{p22}). However, due to the 
divergence indicated by the delta-function in $\sigma_c^{eff}$,
$P_{g+g\to g+c}$ is not computable. The divergence of $\sigma_c^{eff}$ 
corresponds to $s=0$, in which case the momenta of the two incoming gluons are
parallel. In other words, only if $s=0$, a condensate gluon can be produced 
in a $g+g\to g+c$ process. The probability for this process is infinity. The 
two extreme values, zero phase space and infinite collision probability, give 
nevertheless a finite collision rate, as indicated in Eq. (\ref{condrate}). 
However, in numerical calculations, it is almost
impossible to find two particles with parallel momentum. Moreover, it is also
impossible to deal with processes with infinite collision probability.
To overcome this difficulty we have to make an approximation. We define
an energy cutoff $\varepsilon$. Gluons with energy smaller than $\varepsilon$
are regarded as condensate gluons. This approximation, which has the same
mean as the regularization by a nonzero but small effective gluon mass,
breaks the rule that momenta of two incoming gluons in a $g+g\to g+c$ process
should be parallel, or, $s=0$. A $g+g\to g+c$ process is now allowed to occur
with a nonzero but small angle $\alpha_c$ between the momenta of two
incoming gluons, or, with a nonzero but small $s$. Accordingly the divergence
in $\sigma_c^{eff}$ is eliminated, although $\sigma_c^{eff}$ is still large.
The smaller the value of $\varepsilon$, the smaller is $\alpha_c$ and the 
larger is $\sigma_c^{eff}$. Mathematically, the approximation leads to the
replacement of the delta-function in $f^c_1$ by a step function,
\begin{equation}
\label{approx}
\delta^{(3)}({\mathbf p_1}) \approx 
\frac{\theta(\varepsilon-p_1)}{4\pi p_1^2 \varepsilon} \,.
\end{equation}
Putting Eq. (\ref{approx}) into Eq. (\ref{cs22effc}) we obtain
\begin{eqnarray}
\label{cs22effc1}
\sigma_c^{eff}&=&\pi n_c \left[ \frac{| {\cal M}_{34\to 12} |^2}{s} 
\right]_{s=0} \frac{1}{2p} [1+f^g(p)] \nonumber \\
&& \times \frac{1}{4\varepsilon} \left ( \frac{2}{E-p} - \frac{1}{\Delta} 
\right ) \theta \left ( \varepsilon - \frac{E-p}{2} \right ) \,,
\end{eqnarray}
where $\Delta=\min \{\varepsilon, (E+p)/2\}$. In Appendix \ref{appB} one can
find more detailed calculations. Since $s=E^2-p^2$, the step function in 
$\sigma_c^{eff}$ leads to the maximum of $s$, $s_m=2\varepsilon (E+p)$, below 
which a $g+g\to g+c$ can occur. In addition, $s_m$ corresponds to the maximal 
angle between the momenta of two incoming gluons.

In the previous section we have presented the numerical implementation of
$g+g\to g+g$ processes in absence of the gluon condensate. Numerically we
turn off $g+g \to g+c$ processes by setting $\sigma_c^{eff}=0$. On the other 
hand, with the cross section (\ref{cs22no}), collisions can still occur at 
$s<s_m$, and gluons with energy being less than $\varepsilon$ can still be 
produced. Therefore, at the time $t_c$ when we turn on $g+g \to g+c$ processes
to describe the condensation, we will have a nonzero density of condensate 
particles. This density $n_c$ at $t_c$, which is needed to solve 
Eq. (\ref{boltzmann2}), is not physically motivated, but regarded as an 
initial seed for the growth of the condensate. The smaller the value of 
$\varepsilon$, the smaller is $n_c(t_c)$. This will not lead to significant 
difference in the increase of $n_c$, provided that $n_c(t_c)$ is much smaller
than its final equilibrium value, which is true in our case.

With the approximated $\sigma_c^{eff}$ we compute the collision probability
$P_{g+g \to g+c}$ according to Eq. (\ref{p22}) and simulate the $g+g\to g+c$
process. The effect on the collision rate due to the approximation with the 
energy cutoff is negligible, if $\varepsilon$ is small enough. In the present
calculation we set $\varepsilon = 2.5 \mbox{ MeV}$, which corresponds to
the limitation that the extraction of $f$ below $5 \mbox{ MeV}$ is inaccurate
due to small numbers of test particles. In Appendix \ref{appC} we
show the potential moderate effect on the collision rate, if $\varepsilon$
becomes large.

The numerical implementation for back reactions $c+g \to g+g$ is same as
that for $g+g\to g+g$, which has been presented in the previous section. 
Compared with $\sigma_{g+g\to g+g}$ (\ref{cs22no}), the total cross section 
for $c+g\to g+g$ is
\begin{eqnarray}
\sigma_c&=&\frac{1}{2} \int_{-s}^0 dt 
\frac{| {\cal M}_{34\to 12} |^2}{16\pi s^2} \nonumber \\
&=&\frac{1}{32\pi s} \int_{-s}^0 dt 
\left[ \frac{| {\cal M}_{34\to 12} |^2}{s} \right]_{s=0} \nonumber \\
&=&\frac{1}{32\pi s} \int_{-s}^0 dt 144 \pi^2 \alpha_s^2 \frac{1}{m^2_D}
=\frac{9}{2} \pi \frac{\alpha_s^2}{m_D^2} \,.
\end{eqnarray}
Due to the kinematic reason it is always true that $s \le s_m$ in each 
$c+g\to g+g$ process. Numerically, if $s \le s_m$ we use $\sigma_c$, 
if $s > s_m$ we use $\sigma_{g+g\to g+g}$ (\ref{cs22no}). We assume 
a continuous change of the cross section with respect to $s$. Thus, 
$\sigma_{g+g\to g+g}$ should be equal to $\sigma_c$ at $s_m$. With this 
condition we determine $t_{cut}$ in Eq. (\ref{cs22no}), which is 
$t_{cut}=-m_D^2/[e(1+m_D^2/s_m)-1]$.

From Eq. (\ref{boltzmann2}) we recognize that both the collision rate of
$g+g \to g+c$ and that of $c+g \to g+g$ contain a same contribution, which
is proportional to $f_4^g f_3^g f_2^g f_1^c$. Therefore, the term being 
proportional to $f^g(p)=f_2^g$ in Eq. (\ref{cs22effc1}) corresponds to
this contribution in $g+g \to g+c$ processes ($4+3\to 2+1$), while the term
being proportional to $f_3^g f_4^g$ in $P'_{22}$ [see Eq. (\ref{p22p})]
\footnotemark \footnotetext{For a $1+2\to 3+4$ process $f_1$ and $f_2$ in
Eq. (\ref{p22p}) should be replaced by $f_3$ and $f_4$.}
corresponds to the same contribution in $c+g \to g+g$ ($1+2\to 3+4$)
processes. The two contributions cancel out. In numerical calculations we thus
replace $1+f^g(p)$ in Eq. (\ref{cs22effc1}) by $1$ and replace the Bose factor
$(1+f_3^g)(1+f_4^g)$ in $P'_{22}$ by $1+f^g_3+f^g_4$. More details can be
found in Appendixes \ref{appA} and \ref{appB}.

In Fig. \ref{fig10-distbec} we show the time evolution of the momentum 
distribution of gluons from $t_c$, when the condensation begins, to a later 
time $6.692 \mbox{ fm/c}$, when the condensation is complete. 
\begin{figure}[ht]
 \centering
 \includegraphics[width=0.45\textwidth]{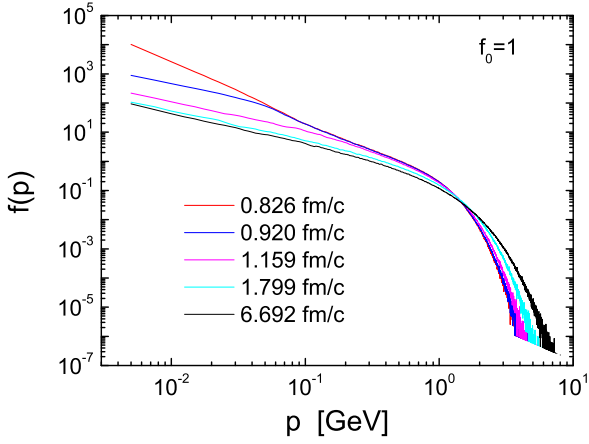}
 \caption{Same as Fig. \ref{fig6-dist}.}
 \label{fig10-distbec}
\end{figure}
With the growing gluon condensate we find a rapid change of the distribution 
at low momentum from the $p^{-2}$ scaling at $t_c=0.826 \mbox{ fm/c}$ to 
$p^{-1}$ at $0.92 \mbox { fm/c}$. During this time the $p^{-7/3}$ scaling 
remains. As the time further proceeds, the $p^{-1}$ scaling extends to larger 
momentum region, so that the region with the $p^{-7/3}$ scaling completely 
disappears at $1.159 \mbox{ fm/c}$.
Due to the growth of the condensate the particle distribution at low momentum
decreases. The energies freed from condensation are transferred to particles
with larger momentum because of the energy conservation. This energy transfer 
leads to the increase of the distribution function at large momentum. 
Figure \ref{fig11-be2} shows the comparison of the gluon distribution at 
$6.692 \mbox{ fm/c}$ with the Bose-Einstein distribution function with
$T=0.443 \mbox{ GeV}$ and $\mu_{eq}=0$.
\begin{figure}[ht]
 \centering
 \includegraphics[width=0.45\textwidth]{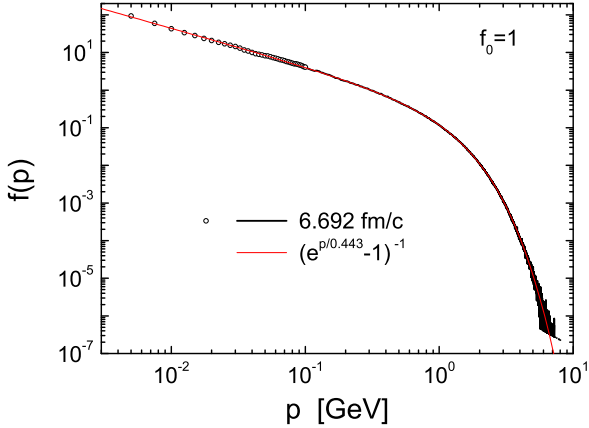}
 \caption{Comparison of the gluon momentum distribution
at $6.692 \mbox{ fm/c}$ (black open circles and solid curve) with the 
Bose-Einstein distribution at thermal equilibrium (red curve).}
 \label{fig11-be2}
\end{figure}
The latter is the thermal equilibrium distribution when the condensation
is complete, see Eq. (\ref{temp}). The open circles depict the first $40$ 
values of the calculated distribution separated by a equidistant interval of 
$\Delta p=2.5 \mbox{ MeV}$ and beginning from $5 \mbox{ MeV}$.
We see that the calculated result agrees nicely with the analytical distribution
over $8$ orders in magnitude. Particularly we see agreements at very low
as well as very high momentum, where strong statistical fluctuations are
expected due to small particle numbers. 

\begin{figure}[ht]
 \centering
 \includegraphics[width=0.45\textwidth]{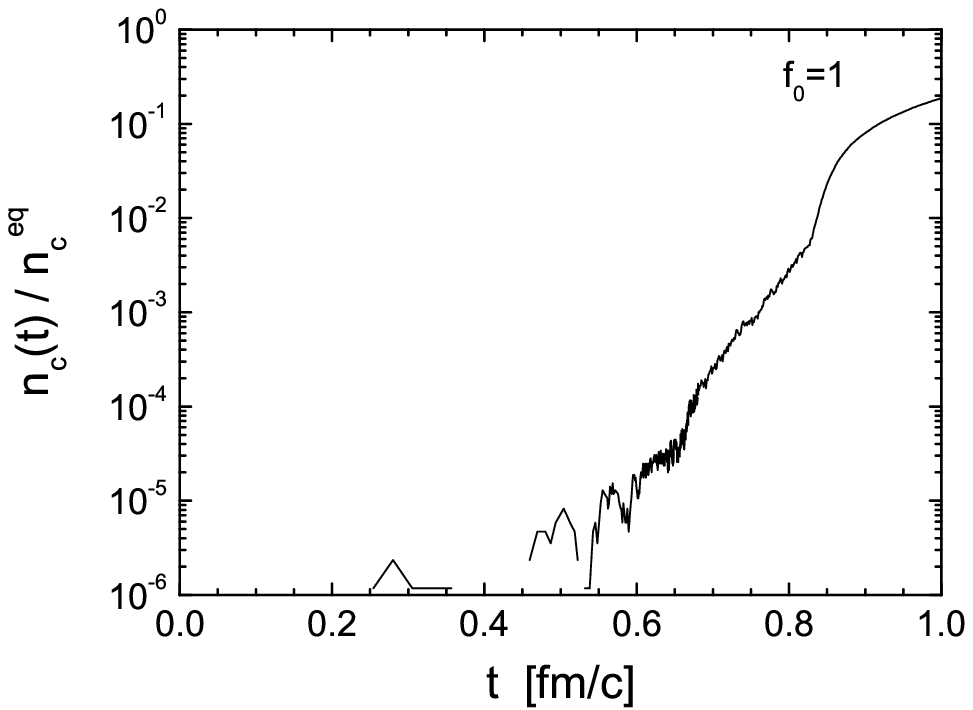}
  \includegraphics[width=0.45\textwidth]{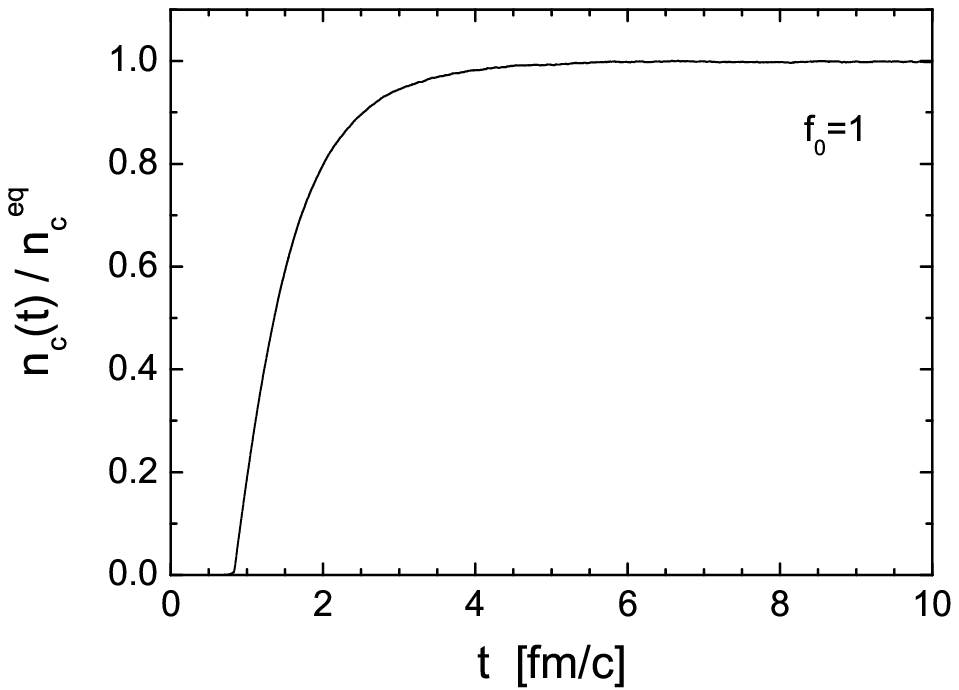}
 \caption{Time evolution of the density of the gluon condensate.}
 \label{fig12-nc}
\end{figure}

We show in Fig. \ref{fig12-nc} the growth of the gluon condensate in time, 
divided by the expected density at equilibrium, given in Eq. (\ref{nceq}).
Before the condensation really starts at $t_c=0.826 \mbox{ fm/c}$, the 
meaning of $n_c$ is the density of gluons, which energy is smaller than 
$2.5 \mbox{ MeV}$. The increase of this density is due to collisions of gluons 
without the presence of a gluon condensate. We see that the increase of this
density is almost exponential. At $t_c$ the density reaches about $1\%$ of 
the value of the condensate density at equilibrium. This value is regarded 
as a seed for the growth of the condensate.

Once the condensation starts, $n_c$ denotes the density of the condensate,
although $n_c$ is still calculated as the density of gluons with energy being
smaller than $2.5 \mbox{ MeV}$ due to the numerical handling explained before
in this section. In the upper panel of Fig. \ref{fig12-nc} we see a much stronger
increase of $n_c$ after $t_c$ than that before $t_c$. At early times of the
condensation the production processes are dominant compared to the evaporation
processes. At these times the gluon condensate grows exponentially, which can 
qualitatively be understood by Eq. (\ref{condrate}). When the evaporation 
processes begin to balance the production processes, the growth of the gluon 
condensate slows down, and $n_c$ relaxes to its final value at thermal 
equilibrium. The relaxation of the calculated $n_c$ to the expected value at 
equilibrium (see the lower panel of Fig. \ref{fig12-nc}) and the agreement of
the distribution of non-condensate gluons with the expected Bose-Einstein
function (see Fig. \ref{fig11-be2}) demonstrate the correct numerical
implementations for solving the Boltzmann equation with the presence of
a Bose-Einstein condensate.

\begin{figure}[ht]
 \centering
 \includegraphics[width=0.45\textwidth]{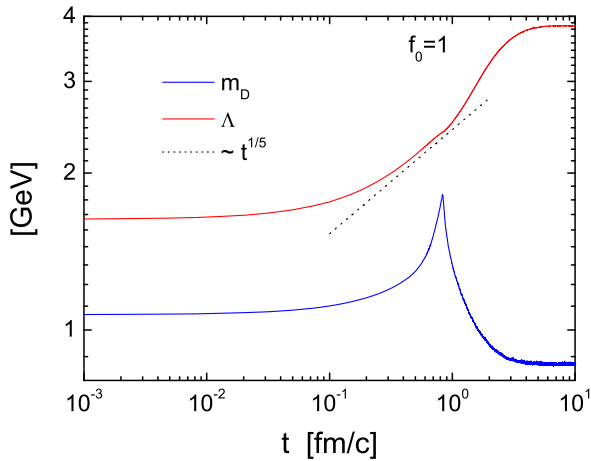}
 \caption{Time evolution of the hard scale $\Lambda$ and the Debye screening
 mass $m_D$.}
 \label{fig13-mdlam}
\end{figure}

During the thermalization the typical hard momentum, which is $Q_s$ initially,
increases, as the energy flows towards the ultraviolet momentum region. 
As suggested in Ref. \cite{Berges:2013fga}, we define the hard momentum
scale $\Lambda(t)$ as
\begin{equation}
\Lambda^2(t)= \frac{\int d\Gamma 4p^2 E f^g({\mathbf p},t)}
{\int d\Gamma E f^g({\mathbf p},t)}\,.
\end{equation}
If the solution of the Boltzmann equation (\ref{boltzmann1}) is 
self-similar, i.e., $f^g({\mathbf p},t)=t^\alpha f_s(t^\beta {\mathbf p})$,
then the hard scale shows a scaling behavior \cite{Berges:2013fga},
$\Lambda(t) \sim t^{-\beta}$. Following the derivation in 
\cite{Micha:2004bv,Berges:2013fga},
the exponents $\alpha$ and $\beta$ can be obtained by putting the self-similar
solution into Eq. (\ref{boltzmann1}) and using the energy conservation. 
The values of the exponents depend on the matrix element.
For our case [see Eq. (\ref{matrix})] we obtain $\alpha=-4/5$ and $\beta=-1/5$.
Figure \ref{fig13-mdlam} shows the time evolution of the hard scale 
$\Lambda(t)$ compared with a function $\sim t^{-1/5}$.
We see the agreement in a time window between $0.3 \mbox{ fm/c}$ and 
$t_c=0.826 \mbox{ fm/c}$.
This indicates that within this time window a self-similar distribution
is achieved over a momentum region, which is sensitive to the hard scale.
After $t_c$ the condensation occurs, which accelerates the energy transportation
towards the ultraviolet momentum region. The increase of $\Lambda(t)$ 
becomes stronger. As the condensation completes, $\Lambda(t)$ relaxes to
its value at equilibrium, which is $4 \pi \sqrt{10/21} T$.

In Fig. \ref{fig13-mdlam} we also show the time evolution of the Debye 
screening mass $m_D$, which squared is defined in Eq. (\ref{md2}) with $f^g$ 
instead of $f$. $m_D(t)$ is governed by low momenta. Before $t_c$
the over-population of the low momentum gluons leads to a rapid increase 
of $m_D$ to infinity. The Bose-Einstein condensation after $t_c$ reduces
the over-population, which then leads to the decrease of $m_D$. 
At thermal equilibrium $m_D$ relaxes to $\sqrt{4\pi\alpha_s} T$.
Both relaxation values of $\Lambda$ and $m_D$ from the calculation agree 
well with the expected equilibrium values.

\section{Summary}
\label{sec5:sum}
In this paper we have presented a new numerical method, which solves Boltzmann
equations for bosons, in particular, with the presence of a Bose-Einstein condensate. 
Compared to the old method, which has been developed in BAMPS to describe collisions
of Boltzmann particles, the new method takes Bose statistics into account by
considering the angular differential collision probability instead of the total
collision probability, which is more time consuming in practice. Moreover, the new
method does not require any approximations to matrix elements of interactions and,
thus, is a general scheme. The numerical implementation of this new method has been
well tested by performing box calculations for static particle systems. First we
have considered systems at thermal equilibrium with
the Bose-Einstein distribution at various temperatures and calculated the collision
rate. We have seen that the calculated collision rates agree well with the expected
analytical values. Second, we have assumed two different non-thermal initial
conditions and evolved systems to the equilibrium states. The expected equilibrium
distributions are Bose-Einstein functions with negative and zero chemical potentials.
We have found that in both cases the final distributions from the calculations
agree well with the expected analytical functions. These successes demonstrate
the correct implementations of Bose statistics in the Boltzmann equation for bosons
through the new method.

Employing the tested numerical implementations we have then investigated the
onset of the Bose-Einstein condensation for an initially over-populated gluon
system, before the Bose-Einstein condensation occurs. Due to the Bose statistical
factor the distribution at low momenta increases quickly to be over-populated and
is thus far from thermal equilibrium. By looking at the time evolution of the momentum
distribution function we have observed the appearance of two power law scalings,
$p^{-2}$ in the infrared and $p^{-7/3}$ in the intermediate momentum region.
The two power law scaling functions have exactly the same exponents as those 
suggested by the scaling arguments for the solutions to the Boltzmann equation in
momentum regions, where $f(p) \gg 1$. The $p^{-7/3}$ scaling function is self-similar
within a small momentum window, which may be sensitive to the hard momentum
scale, because the time evolution of the hard scale shows the expected power 
law scaling behavior before the onset of the Bose-Einstein condensation and 
thus reflects the self-similarity of the particle distribution. The $p^{-2}$
scaling is, however, not yet self-similar as suggested, because
the $p^{-7/3}$ scaling extends to the infrared region, so that the magnitude
of the $p^{-2}$ scaling function increases with time. This behavior leads to 
an energy transport towards $p=0$ besides a particle transport. Since the 
distribution function for $p < 5 \mbox{ MeV}$ cannot be calculated with 
the required accuracy due to the limitation of the current numerical 
computation, our calculation has to be stopped at some time, when the $p^{-2}$
scaling extends to the region of $p < 5 \mbox{ MeV}$. The answer to the 
question whether the $p^{-2}$ scaling will become self-similar at some
later time has to be postponed to a future work. Besides the two scalings
suggested we have also found a further power law scaling, $p^{-7/4}$, 
following the $p^{-7/3}$ scaling. The distribution function in this new scaling 
region is decreasing in time, which leads to the transportation of particles and
energies through the $p^{-7/3}$ scaling region into the $p^{-2}$ scaling region.

Finally the Boltzmann equation is solved with the presence of a Bose-Einstein
condensate. We have found that if the condensate consists of massless 
particles, the matrix element of interactions between condensate and 
non-condensate particles should be constrained by the requirement that the
ratio of the matrix element squared to the invariant mass $s$ must be finite at
$s\to 0$. The matrix element of gluon scatterings, which is motivated by the HTL
calculations and has been already employed in the calculation for the onset of
the gluon condensation, fulfills this constraint. We have continued the calculation
for the onset with a seed for the growth of the condensate and demonstrated the 
gluon condensation from an out of equilibrium state. The condensation reduces 
the over-population of gluons at low momenta. The energy freed from the
condensation is transferred to particles with large momentum. As the condensation
becomes complete, the system of non-condensate gluons approaches thermal
equilibrium, which agrees well with the expected Bose-Einstein distribution.

To know whether a gluon condensate exists and to further understand
thermalization in heavy-ion collisions, we need more further investigations. 
As the next, we will study the role of inelastic scatterings 
\cite{Huang:2013lia,Blaizot:2016iir}
and expansion in the possible formation of a gluon condensate.

\section*{Acknowledgement}
ZX thanks X.G. Huang, J. Liao, R. Venugopalan, and L. McLerran for helpful 
discussions. This work was financially supported by the NSFC and the MOST
under Grants No. 11575092, No. 11335005, No. 2014CB845400, and 
No. 2015CB856903. KZ and CG were supported by  the Helmholtz International
Center for FAIR within the framework of the LOEWE program launched by the
State of Hesse.
The BAMPS simulations were performed at Tsinghua National Laboratory for 
Information Science and Technology.

\appendix

\section{Rate equation of the condensation}
\label{appA}
In this section we derive the rate equation of the condensation, 
Eq. (\ref{condrate}), from Eq. (\ref{boltzmann2}). Integrating 
Eq. (\ref{boltzmann2}) over the momentum of the condensate particle 
$d^3p_1/(2\pi)^3$ gives the time derivative of the density of condensate 
particles
\begin{eqnarray}
\label{app-nc1}
\frac{\partial n_c}{\partial t} &=& \frac{1}{2} \int d\Gamma_1 d\Gamma_2 
\int d\Gamma_3 d\Gamma_4
|{\cal M}_{34\to 12}|^2(2\pi)^4 \nonumber \\
&&\times \ \delta^{(4)} (p_3+p_4-p_1-p_2)  \left[ f_3^{g} f_4^{g} f_1^{c} 
(1+f_2^{g}) \right. \nonumber \\
&& \left.- f_1^{c} f_2^{g} (1+f_3^{g}) (1+f_4^{g}) \right] \nonumber \\
&\equiv& R_c^{gain} - R_c^{loss}\,,
\end{eqnarray}
where the first and second term are named as $R_c^{gain}$ and $R_c^{loss}$ 
denoting the condensation and evaporation rate, respectively.

In the following we carry out integrations in $R_c^{gain}$ explicitly. 
At first we integrate over $d^3p_2$ with help of the delta-function 
$\delta^{(3)}({\mathbf p}_3+{\mathbf p}_4-{\mathbf p}_1-{\mathbf p}_2)$ 
and obtain
\begin{eqnarray}
\label{app-nc2}
R_c^{gain} &=& \frac{1}{2} \int \frac{d^3 p_3}{(2\pi)^3 2E_3} \frac{d^3 p_4}{(2\pi)^3 2E_4}
\frac{d^3 p_1}{(2\pi)^3 2E_1} \frac{1}{2(E-E_1)} \nonumber \\
&& \times \ |{\cal M}_{34\to 12}|^2 
2\pi \delta[F({\mathbf p}_1)] f_3^g f_4^g f_1^c (1+f_2^g) \,,
\end{eqnarray}
where $f_2^g=f^g(E-E_1,{\mathbf p}-{\mathbf p}_1;t)$, $E=E_3+E_4=E_1+E_2$ 
is the total energy and 
${\mathbf p}={\mathbf p}_3+{\mathbf p}_4={\mathbf p}_1+{\mathbf p}_2$ is the
total momentum. $\delta[F({\mathbf p}_1)]$ indicates the energy conservation,
where
\begin{eqnarray}
\label{app-fp}
F({\mathbf p}_1)&=&E-E_1-E_2=E-E_1-\sqrt{p^2_2+m^2} \nonumber \\
&=& E-E_1-\sqrt{({\mathbf p}-{\mathbf p}_1)^2+m^2} \,.
\end{eqnarray}
Using the identity
\begin{equation}
 \label{iden}
\int  dE_1 d^3 p_1 \delta(E_1^2-p_1^2-m^2)= \int \frac{d^3 p_1}{2E_1}
\end{equation}
and $f_1^c=(2\pi)^3 n_c \delta^{(3)} ({\mathbf p}_1)$ we then rewrite Eq. (\ref{app-nc2}) to
\begin{eqnarray}
\label{app-nc3}
R_c^{gain} &=& \pi n_c \int \frac{d^3 p_3}{(2\pi)^3 2E_3} 
\frac{d^3 p_4}{(2\pi)^3 2E_4} f_3^g f_4^g \int dE_1 d^3 p_1  \nonumber \\
&& \times \ \frac{1}{2(E-E_1)} |{\cal M}_{34\to 12}|^2 \delta(E_1^2-p_1^2-m^2)
\delta[F({\mathbf p}_1)] \nonumber \\
&& \times \ \delta^{(3)}({\mathbf p}_1) (1+f_2^g) \,.
\end{eqnarray}

As the next we integrate over $d^3 p_1$ and then $dE_1$ using the delta 
function $\delta^{(3)}({\mathbf p}_1)$ and $\delta[F({\mathbf p}_1)]$
\begin{eqnarray}
\label{app-nc4}
R_c^{gain} &=& \pi n_c \int \frac{d^3 p_3}{(2\pi)^3 2E_3} \frac{d^3 p_4}{(2\pi)^3 2E_4}
f_3^g f_4^g \int d E_1 \frac{1}{2(E-E_1)} \nonumber \\
&& \times \ |{\cal M}_{34\to 12}|^2 \delta(E_1^2-m^2) \delta(E-E_1-\sqrt{p^2+m^2})
\nonumber \\
&& \times \ (1+f_2^g) \nonumber \\
&=& \pi n_c \int \frac{d^3 p_3}{(2\pi)^3 2E_3} \frac{d^3 p_4}{(2\pi)^3 2E_4}
f_3^g f_4^g \frac{1}{2\sqrt{p^2+m^2}} \nonumber \\
&& \times \  |{\cal M}_{34\to 12}|^2 \delta[(E-\sqrt{p^2+m^2})^2-m^2] \nonumber \\
&& \times \ (1+f_2^g)\,,
%  =&& \frac{2\pi n_c}{16(2\pi)^6} \int dp_3 dp_4\frac{p_3^2p_4^2}{E_3 E_4} f_3 f_4 \int d(\cos\theta_3)d(\cos\theta_4)d\phi_3d\phi_4 [1+f_2({\mathbf p})] \nonumber \\
%  &&\times |{\cal M}_{34\to 12}|^2 \frac{1}{\sqrt{{\mathbf p}^2+m^2}} \delta((E-\sqrt{p^2+m^2})^2-m^2)\,.
\end{eqnarray}
where $f_2^g=f^g(\sqrt{p^2+m^2},{\mathbf p};t)$. We denote that $\theta$ is
the angle between ${\mathbf p}_3$ and ${\mathbf p}_4$. Then we have
\begin{equation}
\label{momp}
p=|{\mathbf p}_3+{\mathbf p}_4|=\sqrt{p_3^2+p_4^2+2p_3p_4\cos \theta} \,.
\end{equation}
We assume that the distribution function $f$ is isotropic in momentum space.
Therefore, $f_3^g=f^g(p_3,t)$, $f_4^g=f^g(p_4,t)$, $f_2^g=f^g(p,t)$, and 
we can integrate Eq. (\ref{app-nc4}) over the solid angles of 
${\mathbf p}_3$ and ${\mathbf p}_4$
\begin{eqnarray}
 \label{app-nc5}
R_c^{gain} &=& \frac{n_c}{64\pi^3} \int dp_3 dp_4\frac{p_3^2p_4^2}{E_3 E_4}
f_3^g f_4^g \int d\cos\theta \frac{1}{\sqrt{p^2+m^2}} \nonumber \\
&& \times \  |{\cal M}_{34\to 12}|^2 \delta[(E-\sqrt{p^2+m^2})^2-m^2] 
\nonumber \\
&& \times \ (1+f_2^g) \,.
\end{eqnarray}
The integral over $\cos \theta$ can be carried out using the delta funtion
and gives
\begin{eqnarray}
 \label{app-nc6}
R_c^{gain} &=& \frac{n_c}{64\pi^3} \int dp_3 dp_4\frac{p_3 p_4}{E_3 E_4}
f_3^g f_4^g (1+f_2^g) \frac{1}{2m}  \nonumber \\
&& \times \ \left [ |{\cal M}|^2 \right]_{E-\sqrt{p^2+m^2}=m} \nonumber \\
&=&\frac{n_c}{64\pi^3} \int dE_3 dE_4 f_3^g f_4^g (1+f_2^g) \frac{1}{2m}
\nonumber \\
&& \times \ \left [ |{\cal M}|^2 \right]_{E-\sqrt{p^2+m^2}=m}\,.
\end{eqnarray}
Since $s=E^2-p^2$, the constraint $E-\sqrt{p^2+m^2}=m$ is equivalent to 
$s=2mE$. This leads to 
\begin{equation}
 \label{app-nc7}
R_c^{gain} =\frac{n_c}{64\pi^3} \int dE_3 dE_4 f_3^g f_4^g (1+f_2^g) E
\left [ \frac{|{\cal M}|^2}{s} \right]_{s=2mE}\,,
\end{equation}
where $f_2^g=f^g(p,t)=f^g(\sqrt{E^2-2mE},t)$.

The integrals in $R_c^{loss}$ proceed similar as those shown above.
We obtain $R_c^{loss}$ by replacing $f_4^g f_3^g (1+f_2^g)$ in 
Eq. (\ref{app-nc7}) with $f_2^g(1+f_3^g)(1+f_4^g)$. We have finally
\begin{eqnarray}
 \label{app-nc8}
\frac{\partial n_c}{\partial t} &=&\frac{n_c}{64\pi^3} \int dE_3 dE_4
[f_3^g f_4^g (1+f_2^g) \nonumber \\
&& \ \ \ \ \ \ \ \ \ -(1+f_3^g)(1+f_4^g)f_2^g] E
\left [ \frac{|{\cal M}|^2}{s} \right]_{s=2mE}
\nonumber \\
&=&\frac{n_c}{64\pi^3} \int dE_3 dE_4
[f_3^g f_4^g -f_2^g(1+f_3^g+f_4^g)] \nonumber \\
&& \ \ \ \ \ \ \ \ \ \times E
\left [ \frac{|{\cal M}|^2}{s} \right]_{s=2mE} \,,
\end{eqnarray}
which is Eq. (\ref{condrate}). We see that both $R_c^{gain}$ and $R_c^{loss}$
contain a same contribution, which is proportional to $f_4^g f_3^g f_2^g$.
The two contributions cancel out.

\section{The effective cross section for $g+g\to g+c$}
\label{appB}
According to the definition (\ref{cs22eff}) the effective cross section of
a $g+g\to g+c$ process is
\begin{eqnarray}
\label{app-cs22effc}
\sigma_c^{eff}&=& \frac{1}{2s} \int d\Gamma_1 d\Gamma_2
| {\cal M}_{34\to 12} |^2 f^c_1(1+f^g_2) \nonumber \\
&& \times (2\pi)^4 \delta^{(4)} (p_3+p_4-p_1-p_2) \,.
\end{eqnarray}
Compared with Eqs. (\ref{app-nc1}) and (\ref{app-nc4}), we realize that
\begin{equation}
\label{app-cs22eff1}
R_c^{gain}=\int d\Gamma_3 d\Gamma_4 f_3^g f_4^g s \sigma_c^{eff} \,.
\end{equation}
We thus obtain $\sigma_c^{eff}$, as expressed in Eq. (\ref{cs22effc}).

With the approximation
\begin{equation}
\delta^{(3)}({\mathbf p_1}) \approx 
\frac{\theta(\varepsilon-p_1)}{4\pi p_1^2 \varepsilon}
\end{equation}
Eq. (\ref{app-cs22effc}) is changed to
\begin{eqnarray}
\sigma_c^{eff}&=& \frac{1}{2}  \left [ \frac{| {\cal M}_{34\to 12} |^2}{s}
\right ]_{s=0} [1+f^g(p)] \int d\Gamma_1 d\Gamma_2 \nonumber \\
&& \times (2\pi)^3 n_c \frac{\theta(\varepsilon-p_1)}{4\pi p_1^2 \varepsilon}
(2\pi)^4 \delta^{(4)} (p_3+p_4-p_1-p_2) \,. \nonumber \\
&&
\end{eqnarray}
The integral over $d^3p_2$ with help of the delta-function 
$\delta^{(3)}({\mathbf p}_3+{\mathbf p}_4-{\mathbf p}_1-{\mathbf p}_2)$ gives
\begin{eqnarray}
\sigma_c^{eff}&=& \pi  \left [ \frac{| {\cal M}_{34\to 12} |^2}{s}
\right ]_{s=0} [1+f^g(p)] \int \frac{d^3p_1}{2E_1} \nonumber \\
&& \times \frac{1}{2(E-E_1)} n_c
\frac{\theta(\varepsilon-p_1)}{4\pi p_1^2 \varepsilon}
\delta[F({\mathbf p}_1)] \,.
\end{eqnarray}
$F({\mathbf p}_1)$ is same as Eq. (\ref{app-fp}),
\begin{equation}
F({\mathbf p}_1)=E-p_1-\sqrt{p^2+p_1^2-2pp_1\cos\theta_1} \,,
\end{equation}
where $\theta_1$ is the angle between ${\mathbf p}$ and ${\mathbf p}_1$.
The solution of $F({\mathbf p}_1)=0$ is $p_1=s/2/(E-p\cos\theta_1)$. Because of
$-1 \le \cos\theta_1 \le 1$, we obtain the limitations for $p_1$,
$(E-p)/2 \le p_1 \le (E+p)/2$.
Without loss of generality ${\mathbf p}$ lies in the $Z$ direction. 
We carry out the integral over the solid angle of ${\mathbf p}_1$ and obtain
\begin{eqnarray}
\sigma_c^{eff}&=& \pi  n_c \left [ \frac{| {\cal M}_{34\to 12} |^2}{s}
\right ]_{s=0} [1+f^g(p)] \int_{(E-p)/2}^{(E+p)/2} dp_1 \nonumber \\
&& \times \frac{1}{4p_1(E-p_1)} \frac{\theta(\varepsilon-p_1)}{2 \varepsilon}
\int_{-1}^1 d\cos\theta_1 \delta[F(p_1,\cos\theta_1)] \nonumber \\
&=& \pi  n_c \left [ \frac{| {\cal M}_{34\to 12} |^2}{s}
\right ]_{s=0} [1+f^g(p)] \int_{(E-p)/2}^{(E+p)/2} dp_1 \nonumber \\
&& \times \frac{1}{4p_1(E-p_1)} \frac{\theta(\varepsilon-p_1)}{2 \varepsilon}
\frac{E-p_1}{pp_1} \nonumber \\
&=& \pi  n_c \left [ \frac{| {\cal M}_{34\to 12} |^2}{s}
\right ]_{s=0} [1+f^g(p)] \int_{(E-p)/2}^{(E+p)/2} dp_1 \nonumber \\
&& \times \frac{1}{4pp_1^2} \frac{\theta(\varepsilon-p_1)}{2 \varepsilon}\,.
\end{eqnarray}
Due to the step function $\sigma_c^{eff}$ is nonzero, only if the lower 
limit $(E-p)/2$ is smaller than $\varepsilon$. The upper limit is changed to
the minimum of $(E+p)/2$ and $\varepsilon$, denoted by $\Delta$.
Integral over $p_1$ gives
\begin{eqnarray}
\label{app-cs22eff2}
\sigma_c^{eff}&=&\pi n_c \left[ \frac{| {\cal M}_{34\to 12} |^2}{s} 
\right]_{s=0} \frac{1}{2p} [1+f^g(p)] \nonumber \\
&& \times \frac{1}{4\varepsilon} \left ( \frac{2}{E-p} - \frac{1}{\Delta} 
\right ) \theta \left ( \varepsilon - \frac{E-p}{2} \right ) \,,
\end{eqnarray}
which is Eq. (\ref{cs22effc1}). From the derivation of
$\partial n_c/\partial t$ [see Eq. (\ref{app-nc8})] we realize that
the term being proportional to $f_2^g f_3^g f_4^g$ in both $R_c^{gain}$
and $R_c^{loss}$ term cancel out. Therefore, in numerical calculations
we replace $[1+f^g(p)]$ in Eq. (\ref{app-cs22eff2}) by $1$,
\begin{eqnarray}
\label{app-cs22eff3}
\sigma_c^{eff}&=&\pi n_c \left[ \frac{| {\cal M}_{34\to 12} |^2}{s} 
\right]_{s=0} \frac{1}{2p} \frac{1}{4\varepsilon} \left ( 
\frac{2}{E-p} - \frac{1}{\Delta} \right ) \nonumber \\
&& \times \theta \left ( \varepsilon - \frac{E-p}{2} \right ) \,,
\end{eqnarray}
since $\sigma_c^{eff}$ relates to $R_c^{gain}$ according to 
Eq. (\ref{app-cs22eff1}).
Accordingly, the Bose factor $(1+f_3^g)(1+f_4^g)$ in the back reaction 
$c+g\to g+g$ is replaced by $1+f_3^g+f_4^g$.

\section{Dependence of the condensation rate on the energy cutoff 
$\varepsilon$}
\label{appC}
We present the dependence of the condensation rate on the energy cutoff 
$\varepsilon$ used in Eq. (\ref{app-cs22eff3}) for numerical evaluations. 
For this purpose we consider a boson system with the 
presence of a BEC at equilibrium. We would perform cascade calculations using 
BAMPS to extract the condensation rate and compare it to the first term of
Eq. (\ref{app-nc8}). However, putting Bose-Einstein distributions into the first
term of Eq. (\ref{app-nc8}) gives an infinite rate, which is impossible 
to compare. Since we are focusing on the dependence of the condensation rate
on $\varepsilon$, we sample the momentum of bosons with the Boltzmann 
distribution. We then run BAMPS with $\sigma_c^{eff}$ 
[Eq. (\ref{app-cs22eff3})] for $g+g\to g+c$ collisions for just one timestep.
We can still evaluate the collision rate numerically, because the test 
particle number $N_{test}$ is set to be sufficient large.

For simplicity we assume elastic collisions with isotropic collision angles,
which means that $|{\cal M}|^2=32\pi s \sigma_{22}$. For constant $\sigma_{22}$
the condensation rate can be obtained analytically,
\begin{eqnarray}
\tilde R_c^{gain} &=&\frac{n_c}{64\pi^3} \left [ \frac{|{\cal M}|^2}{s} 
\right]_{s=0} \int dE_3 dE_4 f_3^g f_4^g E \nonumber \\
&=&n_c\frac{\sigma_{22}}{2\pi^2}\int dE_3dE_4 e^{-\frac{E_3}{T}}
e^{-\frac{E_4}{T}} (E_3+E_4) \nonumber \\
&=&n_c \frac{\sigma_{22} T^3}{\pi^2}.
\label{app-cal}
\end{eqnarray}
In the calculations we set $\sigma_{22}=16 \mbox{ fm}^2$ and 
$T=0.4 \mbox{ GeV}$, which leads to 
$\tilde R_c^{gain}/n_c=13.5 \mbox{ fm}^{-1}$.
Table \ref{app-table} shows the calculated rates and the comparisons with
the exact one ($13.5 \mbox{ fm}^{-1}$) in dependence of $\varepsilon$. 
``err'' means the relative difference between the numerical and analytical
rate.
\begin{table}[b]
\begin{tabular}{|c|c|c|c|c|}
\hline
$\varepsilon$ (MeV) & $0.1$ & $1$ & $10$ & $100$ \\
\hline
$\tilde R_c^{gain}/n_c [\mbox{fm}^{-1}]$, numerical & 13.53 & 13.75 & 13.7 & 11.5 \\
\hline
err & $0.2\%$ & $1.8\%$ & $1.5\%$ & $14.8\%$ \\
\hline
\end{tabular}
\caption{Dependence of the condensation rate on the energy cutoff.}
\label{app-table}
\end{table}
The numerical error becomes significant for increasing $\varepsilon$. 
For the study presented in the main text we have used 
$\varepsilon=2.5 \mbox{ MeV}$, which has a negligible effect on the
condensation rate.


\begin{thebibliography}{88}
%\cite{McLerran:1993ni}
\bibitem{McLerran:1993ni} 
  L.~D.~McLerran and R.~Venugopalan,
  %``Computing quark and gluon distribution functions for very large nuclei,''
  Phys.\ Rev.\ D {\bf 49}, 2233 (1994)
  [hep-ph/9309289];
  %%CITATION = doi:10.1103/PhysRevD.49.2233;%%
  %1613 citations counted in INSPIRE as of 27 Nov 2016
%\cite{McLerran:1993ka}
%\bibitem{McLerran:1993ka} 
  L.~D.~McLerran and R.~Venugopalan,
  %``Gluon distribution functions for very large nuclei at small transverse momentum,''
  Phys.\ Rev.\ D {\bf 49}, 3352 (1994)
  [hep-ph/9311205];
  %%CITATION = doi:10.1103/PhysRevD.49.3352;%%
  %1172 citations counted in INSPIRE as of 27 Nov 2016
 %\cite{McLerran:1994vd}
%\bibitem{McLerran:1994vd} 
  L.~D.~McLerran and R.~Venugopalan,
  %``Green's functions in the color field of a large nucleus,''
  Phys.\ Rev.\ D {\bf 50}, 2225 (1994)
  [hep-ph/9402335].
  %%CITATION = doi:10.1103/PhysRevD.50.2225;%%
  %845 citations counted in INSPIRE as of 27 Nov 2016

%\cite{Gelis:2010nm}
\bibitem{Gelis:2010nm} 
  F.~Gelis, E.~Iancu, J.~Jalilian-Marian and R.~Venugopalan,
  %``The Color Glass Condensate,''
  Ann.\ Rev.\ Nucl.\ Part.\ Sci.\  {\bf 60}, 463 (2010)
  [arXiv:1002.0333 [hep-ph]].
  %%CITATION = doi:10.1146/annurev.nucl.010909.083629;%%
  %515 citations counted in INSPIRE as of 27 Nov 2016

%\cite{Lappi:2006fp}
\bibitem{Lappi:2006fp} 
  T.~Lappi and L.~McLerran,
  %``Some features of the glasma,''
  Nucl.\ Phys.\ A {\bf 772}, 200 (2006)
  [hep-ph/0602189].
  %%CITATION = doi:10.1016/j.nuclphysa.2006.04.001;%%
  %351 citations counted in INSPIRE as of 27 Nov 2016

%\cite{Weigert:2005us}
\bibitem{Weigert:2005us} 
  H.~Weigert,
  %``Evolution at small x(bj): The Color glass condensate,''
  Prog.\ Part.\ Nucl.\ Phys.\  {\bf 55}, 461 (2005)
  [hep-ph/0501087].
  %%CITATION = doi:10.1016/j.ppnp.2005.01.029;%%
  %262 citations counted in INSPIRE as of 27 Nov 2016

%\cite{Gelis:2013rba}
\bibitem{Gelis:2013rba} 
  T.~Epelbaum and F.~Gelis,
  %``Pressure isotropization in high energy heavy ion collisions,''
  Phys.\ Rev.\ Lett.\  {\bf 111}, 232301 (2013)
  doi:10.1103/PhysRevLett.111.232301
  [arXiv:1307.2214 [hep-ph]].
  %%CITATION = doi:10.1103/PhysRevLett.111.232301;%%
  %112 citations counted in INSPIRE as of 28 Feb 2017

%\cite{Kurkela:2015qoa}
\bibitem{Kurkela:2015qoa} 
  A.~Kurkela and Y.~Zhu,
  %``Isotropization and hydrodynamization in weakly coupled heavy-ion collisions,''
  Phys.\ Rev.\ Lett.\  {\bf 115}, no. 18, 182301 (2015)
  doi:10.1103/PhysRevLett.115.182301
  [arXiv:1506.06647 [hep-ph]].
  %%CITATION = doi:10.1103/PhysRevLett.115.182301;%%
  %47 citations counted in INSPIRE as of 28 Feb 2017

%\cite{Blaizot:2011xf}
\bibitem{Blaizot:2011xf} 
  J.~P.~Blaizot, F.~Gelis, J.~F.~Liao, L.~McLerran and R.~Venugopalan,
  %``Bose--Einstein Condensation and Thermalization of the Quark Gluon Plasma,''
  Nucl.\ Phys.\ A {\bf 873}, 68 (2012)
  [arXiv:1107.5296 [hep-ph]].
  %%CITATION = doi:10.1016/j.nuclphysa.2011.10.005;%%
  %121 citations counted in INSPIRE as of 27 Nov 2016

%\cite{Blaizot:2012qd}
\bibitem{Blaizot:2012qd} 
  J.~P.~Blaizot, F.~Gelis, J.~Liao, L.~McLerran and R.~Venugopalan,
  %``Thermalization and Bose-Einstein Condensation in Overpopulated Glasma,''
  Nucl.\ Phys.\ A {\bf 904-905}, 829c (2013)
  [arXiv:1210.6838 [hep-ph]].
  %%CITATION = doi:10.1016/j.nuclphysa.2013.02.144;%%
  %10 citations counted in INSPIRE as of 27 Nov 2016

%\cite{Blaizot:2013lga}
\bibitem{Blaizot:2013lga} 
  J.~P.~Blaizot, J.~Liao and L.~McLerran,
  %``Gluon Transport Equation in the Small Angle Approximation and the Onset of Bose-Einstein Condensation,''
  Nucl.\ Phys.\ A {\bf 920}, 58 (2013)
  [arXiv:1305.2119 [hep-ph]].
  %%CITATION = doi:10.1016/j.nuclphysa.2013.10.010;%%
  %47 citations counted in INSPIRE as of 27 Nov 2016

%\cite{Blaizot:2014sha}
\bibitem{Blaizot:2014sha} 
  J.~P.~Blaizot, J.~Liao and L.~McLerran,
  %``Gluon transport equation in the small angle approximation and the onset of Bose$(G!9(BEinstein condensation,''
  Nucl.\ Phys.\ A {\bf 931}, 359 (2014).
  %%CITATION = doi:10.1016/j.nuclphysa.2014.10.012;%%

%\cite{Huang:2013lia}
\bibitem{Huang:2013lia} 
  X.~G.~Huang and J.~Liao,
  %``Glasma Evolution and Bose-Einstein Condensation with Elastic and Inelastic Collisions,''
  Phys.\ Rev.\ D {\bf 91}, no. 11, 116012 (2015)
  [arXiv:1303.7214 [nucl-th]].
  %%CITATION = doi:10.1103/PhysRevD.91.116012;%%
  %24 citations counted in INSPIRE as of 27 Nov 2016

%\cite{Blaizot:2014jna}
\bibitem{Blaizot:2014jna} 
  J.~P.~Blaizot, B.~Wu and L.~Yan,
  %``Quark production, Bose$(G!9(BEinstein condensates and thermalization of the quark$(G!9(Bgluon plasma,''
  Nucl.\ Phys.\ A {\bf 930}, 139 (2014)
  [arXiv:1402.5049 [hep-ph]].
  %%CITATION = doi:10.1016/j.nuclphysa.2014.07.041;%%
  %20 citations counted in INSPIRE as of 27 Nov 2016

%\cite{Blaizot:2015wga}
\bibitem{Blaizot:2015wga} 
  J.~P.~Blaizot, Y.~Jiang and J.~Liao,
  %``Gluon transport equation with effective mass and dynamical onset of Bose$(G!9(BEinstein condensation,''
  Nucl.\ Phys.\ A {\bf 949}, 48 (2016)
  [arXiv:1503.07260 [hep-ph]].
  %%CITATION = doi:10.1016/j.nuclphysa.2015.07.021;%%
  %5 citations counted in INSPIRE as of 27 Nov 2016

%\cite{Blaizot:2015xga}
\bibitem{Blaizot:2015xga} 
  J.~P.~Blaizot and J.~Liao,
  %``Gluon Transport Equations with Condensate in the Small Angle Approximation,''
  Nucl.\ Phys.\ A {\bf 949}, 35 (2016)
  [arXiv:1503.07263 [hep-ph]].
  %%CITATION = doi:10.1016/j.nuclphysa.2015.08.004;%%
  %4 citations counted in INSPIRE as of 27 Nov 2016

%\cite{Huang:2014iwa}
\bibitem{Huang:2014iwa} 
  X.~G.~Huang and J.~Liao,
  %``Kinetic evolution of the glasma and thermalization in heavy ion collisions,''
  Int.\ J.\ Mod.\ Phys.\ E {\bf 23}, 1430003 (2014)
  [arXiv:1402.5578 [nucl-th]].
  %%CITATION = doi:10.1142/S0218301314300033;%%
  %16 citations counted in INSPIRE as of 27 Nov 2016

%\cite{Blaizot:2016iir}
\bibitem{Blaizot:2016iir} 
  J.~P.~Blaizot, J.~Liao and Y.~Mehtar-Tani,
  %``The thermalization of soft modes in non-expanding isotropic quark gluon plasmas,''
  arXiv:1609.02580 [hep-ph].
  %%CITATION = ARXIV:1609.02580;%%
  %1 citations counted in INSPIRE as of 27 Nov 2016

%\cite{Berges:2013eia}
\bibitem{Berges:2013eia} 
  J.~Berges, K.~Boguslavski, S.~Schlichting and R.~Venugopalan,
  %``Turbulent thermalization process in heavy-ion collisions at ultrarelativistic energies,''
  Phys.\ Rev.\ D {\bf 89}, no. 7, 074011 (2014)
  [arXiv:1303.5650 [hep-ph]].
  %%CITATION = doi:10.1103/PhysRevD.89.074011;%%
  %100 citations counted in INSPIRE as of 27 Nov 2016

%\cite{Berges:2013fga}
\bibitem{Berges:2013fga} 
  J.~Berges, K.~Boguslavski, S.~Schlichting and R.~Venugopalan,
  %``Universal attractor in a highly occupied non-Abelian plasma,''
  Phys.\ Rev.\ D {\bf 89}, no. 11, 114007 (2014)
  [arXiv:1311.3005 [hep-ph]].
  %%CITATION = doi:10.1103/PhysRevD.89.114007;%%
  %77 citations counted in INSPIRE as of 27 Nov 2016

%\cite{Kurkela:2012hp}
\bibitem{Kurkela:2012hp} 
  A.~Kurkela and G.~D.~Moore,
  %``UV Cascade in Classical Yang-Mills Theory,''
  Phys.\ Rev.\ D {\bf 86}, 056008 (2012)
  [arXiv:1207.1663 [hep-ph]].
  %%CITATION = doi:10.1103/PhysRevD.86.056008;%%
  %48 citations counted in INSPIRE as of 27 Nov 2016

%\cite{Schenke:2016ksl}
\bibitem{Schenke:2016ksl} 
  B.~Schenke and S.~Schlichting,
  %``3D glasma initial state for relativistic heavy ion collisions,''
  Phys.\ Rev.\ C {\bf 94}, no. 4, 044907 (2016)
  [arXiv:1605.07158 [hep-ph]].
  %%CITATION = doi:10.1103/PhysRevC.94.044907;%%
  %5 citations counted in INSPIRE as of 27 Nov 2016

%\cite{Berges:2012us}
\bibitem{Berges:2012us} 
  J.~Berges and D.~Sexty,
  %``Bose condensation far from equilibrium,''
  Phys.\ Rev.\ Lett.\  {\bf 108}, 161601 (2012)
  [arXiv:1201.0687 [hep-ph]].
  %%CITATION = doi:10.1103/PhysRevLett.108.161601;%%
  %71 citations counted in INSPIRE as of 27 Nov 2016

%\cite{Berges:2010ez}
\bibitem{Berges:2010ez} 
  J.~Berges and D.~Sexty,
  %``Strong versus weak wave-turbulence in relativistic field theory,''
  Phys.\ Rev.\ D {\bf 83}, 085004 (2011)
  [arXiv:1012.5944 [hep-ph]].
  %%CITATION = doi:10.1103/PhysRevD.83.085004;%%
  %33 citations counted in INSPIRE as of 27 Nov 2016

\bibitem{photon} J. Klaers, J. Schmitt, F. Vewinger and M. Weitz, Nature {\bf 468}, 545(2010).

%\cite{Kurkela:2014tea}
\bibitem{Kurkela:2014tea} 
  A.~Kurkela and E.~Lu,
  %``Approach to Equilibrium in Weakly Coupled Non-Abelian Plasmas,''
  Phys.\ Rev.\ Lett.\  {\bf 113}, no. 18, 182301 (2014)
  [arXiv:1405.6318 [hep-ph]].
  %%CITATION = doi:10.1103/PhysRevLett.113.182301;%%
  %42 citations counted in INSPIRE as of 27 Nov 2016

%\cite{Kurkela:2011ti}
\bibitem{Kurkela:2011ti} 
  A.~Kurkela and G.~D.~Moore,
  %``Thermalization in Weakly Coupled Nonabelian Plasmas,''
  JHEP {\bf 1112}, 044 (2011)
  [arXiv:1107.5050 [hep-ph]].
  %%CITATION = doi:10.1007/JHEP12(2011)044;%%
  %108 citations counted in INSPIRE as of 27 Nov 2016

%\cite{Semikoz:1994zp}
\bibitem{Semikoz:1994zp} 
  D.~V.~Semikoz and I.~I.~Tkachev,
  %``Kinetics of Bose condensation,''
  Phys.\ Rev.\ Lett.\  {\bf 74}, 3093 (1995)
  [hep-ph/9409202];
  %%CITATION = doi:10.1103/PhysRevLett.74.3093;%%
  %58 citations counted in INSPIRE as of 27 Nov 2016
%\cite{Semikoz:1995rd}
%\bibitem{Semikoz:1995rd} 
  D.~V.~Semikoz and I.~I.~Tkachev,
  %``Condensation of bosons in kinetic regime,''
  Phys.\ Rev.\ D {\bf 55}, 489 (1997)
  [hep-ph/9507306].
  %%CITATION = doi:10.1103/PhysRevD.55.489;%%
  %46 citations counted in INSPIRE as of 27 Nov 2016
  
%\cite{Lacaze:2001qf}
\bibitem{Lacaze:2001qf} 
  R.~Lacaze, P.~Lallemand, Y.~Pomeau and S.~Rica,
  %``Dynamical formation of a Bose-Einstein condensate,''
  Physica D {\bf 152}, 779 (2001).
  %%CITATION = PHYSA,D152,779;%%
  %5 citations counted in INSPIRE as of 27 Nov 2016

%\cite{Pantel:2012pi}
\bibitem{Pantel:2012pi} 
  P.~A.~Pantel, D.~Davesne, S.~Chiacchiera and M.~Urban,
  %``Trap anharmonicity and sloshing mode of a Fermi gas,''
  Phys.\ Rev.\ A {\bf 86}, 023635 (2012)
  [arXiv:1206.5688 [cond-mat.quant-gas]].
  %%CITATION = doi:10.1103/PhysRevA.86.023635;%%
  %2 citations counted in INSPIRE as of 27 Nov 2016

\bibitem{pomeau} C. Connaughton and Y. Pomeau, C. R. Physique {\bf 5}, 91-106(2004).

%\cite{jackson}
\bibitem{jackson} 
  B.~Jackson and E.~Zaremba,
  %``Modeling Bose-Einstein condensed gases at finite temperature with N-body simulations,''
  Phys.\ Rev.\ A {\bf 66}, 033606 (2002).

%\cite{Meistrenko:2015mda}
\bibitem{Meistrenko:2015mda} 
  A.~Meistrenko, H.~van Hees, K.~Zhou and C.~Greiner,
  %``Kinetic approach to a relativistic Bose-Einstein condensate,''
  Phys.\ Rev.\ E {\bf 93}, no. 3, 032131 (2016)
  [arXiv:1510.04552 [hep-ph]].
  %%CITATION = doi:10.1103/PhysRevE.93.032131;%%
  %1 citations counted in INSPIRE as of 27 Nov 2016

%\cite{Epelbaum:2015vxa}
\bibitem{Epelbaum:2015vxa} 
  T.~Epelbaum, F.~Gelis, S.~Jeon, G.~Moore and B.~Wu,
  %``Kinetic theory of a longitudinally expanding system of scalar particles,''
  JHEP {\bf 1509}, 117 (2015)
  [arXiv:1506.05580 [hep-ph]].
  %%CITATION = doi:10.1007/JHEP09(2015)117;%%
  %9 citations counted in INSPIRE as of 27 Nov 2016

%\cite{Dusling:2010rm}
\bibitem{Dusling:2010rm} 
  K.~Dusling, T.~Epelbaum, F.~Gelis and R.~Venugopalan,
  %``Role of quantum fluctuations in a system with strong fields: Onset of hydrodynamical flow,''
  Nucl.\ Phys.\ A {\bf 850}, 69 (2011)
  [arXiv:1009.4363 [hep-ph]].
  %%CITATION = doi:10.1016/j.nuclphysa.2010.11.009;%%
  %81 citations counted in INSPIRE as of 27 Nov 2016

%\cite{Epelbaum:2011pc}
\bibitem{Epelbaum:2011pc} 
  T.~Epelbaum and F.~Gelis,
  %``Role of quantum fluctuations in a system with strong fields: Spectral properties and Thermalization,''
  Nucl.\ Phys.\ A {\bf 872}, 210 (2011)
  [arXiv:1107.0668 [hep-ph]].
  %%CITATION = doi:10.1016/j.nuclphysa.2011.09.019;%%
  %64 citations counted in INSPIRE as of 27 Nov 2016
  
%\cite{Berges:2015ixa}
\bibitem{Berges:2015ixa} 
  J.~Berges, K.~Boguslavski, S.~Schlichting and R.~Venugopalan,
  %``Nonequilibrium fixed points in longitudinally expanding scalar theories: infrared cascade, Bose condensation and a challenge for kinetic theory,''
  Phys.\ Rev.\ D {\bf 92}, no. 9, 096006 (2015)
  [arXiv:1508.03073 [hep-ph]].
  %%CITATION = doi:10.1103/PhysRevD.92.096006;%%
  %6 citations counted in INSPIRE as of 27 Nov 2016

%\cite{Berges:2014xea}
\bibitem{Berges:2014xea} 
  J.~Berges and J.~Jaeckel,
  %``Far from equilibrium dynamics of Bose-Einstein condensation for Axion Dark Matter,''
  Phys.\ Rev.\ D {\bf 91}, no. 2, 025020 (2015)
  [arXiv:1402.4776 [hep-ph]].
  %%CITATION = doi:10.1103/PhysRevD.91.025020;%%
  %14 citations counted in INSPIRE as of 27 Nov 2016

%\cite{Xu:2004mz}
\bibitem{Xu:2004mz} 
  Z.~Xu and C.~Greiner,
  %``Thermalization of gluons in ultrarelativistic heavy ion collisions by including three-body interactions in a parton cascade,''
  Phys.\ Rev.\ C {\bf 71}, 064901 (2005)
  [hep-ph/0406278];
  %%CITATION = doi:10.1103/PhysRevC.71.064901;%%
  %347 citations counted in INSPIRE as of 27 Nov 2016
%\cite{Xu:2007aa}
%\bibitem{Xu:2007aa} 
  Z.~Xu and C.~Greiner,
  %``Transport rates and momentum isotropization of gluon matter in ultrarelativistic heavy-ion collisions,''
  Phys.\ Rev.\ C {\bf 76}, 024911 (2007)
  [hep-ph/0703233].
  %%CITATION = doi:10.1103/PhysRevC.76.024911;%%
  %121 citations counted in INSPIRE as of 27 Nov 2016

%\cite{Xu:2014ega}
\bibitem{Xu:2014ega} 
  Z.~Xu, K.~Zhou, P.~Zhuang and C.~Greiner,
  %``Thermalization of gluons with Bose-Einstein condensation,''
  Phys.\ Rev.\ Lett.\  {\bf 114}, no. 18, 182301 (2015)
  [arXiv:1410.5616 [hep-ph]].
  %%CITATION = doi:10.1103/PhysRevLett.114.182301;%%
  %15 citations counted in INSPIRE as of 27 Nov 2016
  
%\cite{Scardina:2014gxa}
\bibitem{Scardina:2014gxa} 
  F.~Scardina, D.~Perricone, S.~Plumari, M.~Ruggieri and V.~Greco,
  %``Relativistic Boltzmann transport approach with Bose-Einstein statistics and the onset of gluon condensation,''
  Phys.\ Rev.\ C {\bf 90}, no. 5, 054904 (2014)
  [arXiv:1408.1313 [nucl-th]].
  %%CITATION = doi:10.1103/PhysRevC.90.054904;%%
  %15 citations counted in INSPIRE as of 27 Nov 2016

%\cite{Aurenche:2002pd}
\bibitem{Aurenche:2002pd} 
  P.~Aurenche, F.~Gelis and H.~Zaraket,
  %``A Simple sum rule for the thermal gluon spectral function and applications,''
  JHEP {\bf 0205}, 043 (2002)
  [hep-ph/0204146].
  %%CITATION = HEP-PH/0204146;%%
  %64 citations counted in INSPIRE as of 05 Oct 2014

%\cite{Micha:2004bv}
\bibitem{Micha:2004bv} 
  R.~Micha and I.~I.~Tkachev,
  %``Turbulent thermalization,''
  Phys.\ Rev.\ D {\bf 70}, 043538 (2004)
  doi:10.1103/PhysRevD.70.043538
  [hep-ph/0403101].
  %%CITATION = doi:10.1103/PhysRevD.70.043538;%%
  %146 citations counted in INSPIRE as of 14 Dec 2016

\end{thebibliography}
\end{document}